\documentclass[11pt]{article}

\usepackage{a4wide}
\usepackage{multirow}
\usepackage{graphicx}
\usepackage{amsmath}
\usepackage{amssymb}
\usepackage[compress]{cite}
\usepackage{hyperref}

\def\beq{\begin{equation}}
\def\eeq{\end{equation}}
\def\be{\begin{equation}}
\def\ee{\end{equation}}
\def\ba{\begin{eqnarray}}
\def\ea{\end{eqnarray}}
\def\FONLL{{\small FONLL}}
\def\HW{{\small HERWIG}}
\def\PW{{\small POWHEG}}
\def\PWPY{{\small POWHEG-PY}}
\def\PWHW{{\small POWHEG-HW}}
\def\MCNLO{{\small MC@NLO}}
\def\HW{{\small HERWIG}}
\def\HWpp{Herwig{\small++}}
\def\PY{{\small PYTHIA}}
\def \lsim{\mathrel{\vcenter
     {\hbox{$<$}\nointerlineskip\hbox{$\sim$}}}}

\newcommand\sss{\scriptscriptstyle\rm}
\newcommand\as{\alpha_{\sss S}}

\title{\textbf{Theoretical predictions for charm and bottom production 
at the LHC}}

\author{Matteo Cacciari$^{a,b}$, Stefano Frixione$^{c,d}$\footnote{On leave 
of absence from INFN, Sezione di Genova, Genoa, Italy.},
Nicolas Houdeau$^a$,\\
Michelangelo L. Mangano$^c$, 
Paolo Nason$^{c,e}$\footnote{On leave 
of absence from INFN, Sezione di Milano Bicocca, Milan, Italy.} 
and Giovanni Ridolfi$^{f}$\\[5pt]
~\\
\small
$^a$ LPTHE, UPMC Universit\'e Paris 6 and CNRS UMR 7589, Paris, France\\[5pt]
\small
$^b$ Universit\'e Paris Diderot -- Paris 7, Paris, France\\[5pt]
\small
$^c$ PH Department, TH Unit, CERN, CH-1211 Geneva 23, Switzerland\\[5pt]
\small
$^d$ Institut de Th\'eorie des Ph\'enom\`enes Physiques, EPFL, 
CH-1015 Lausanne, Switzerland\\[5pt]
\small
$^e$ INFN, Sezione di Milano-Bicocca,\\
\small
Piazza della Scienza 3, 20126 Milan, Italy\\[5pt]
\small
$^f$ Dipartimento di Fisica, Universit\`a di Genova, and INFN, 
Sezione di Genova,\\
\small
Via Dodecaneso 33, I-16146 Genoa, Italy.\\[5pt]
\small
E-mail: matteo.cacciari@lpthe.jussieu.fr, stefano.frixione@cern.ch, \\
\small
nicolas.houdeau@lpthe.jussieu.fr, 
michelangelo.mangano@cern.ch,\\
\small
paolo.nason@mib.infn.it, giovanni.ridolfi@ge.infn.it
}

\begin{document}

\maketitle
\date{}

\vspace{1cm}
\begin{abstract}
We present predictions for a variety of single-inclusive observables
that stem from the production of charm and bottom quark pairs at
the 7~TeV LHC. They are obtained within the \FONLL\ semi-analytical 
framework, and with two ``Monte Carlo + NLO'' approaches, \MCNLO\ and \PW.
Results are given for final states and acceptance cuts that are as close 
as possible to those used by experimental collaborations and, where feasible, 
are compared to LHC data.
\end{abstract}

\vspace{1cm}
\begin{flushright}
CERN-PH-TH/2011-227\\
May 2012
\end{flushright}
\vfill

\section{Introduction\label{sec:intro}}

Measurements of charm and bottom production are among the most
interesting studies of QCD dynamics that have emerged from the
analysis of the first LHC data at 7 TeV centre-of-mass energy. The
large total cross sections, expected to be of the order of 5 mb
(charm) and 250 $\mu$b (bottom), have provided abundant data samples
already with the $\sim 50$~pb$^{-1}$ collected in 2010. A precise
knowledge of charm and bottom quark cross sections and distributions
is important in order to assess the accuracy of QCD calculations, but also
as a test of theoretical predictions susceptible of being used to
estimate backgrounds to new physics searches. This is all the more
true since, in the past, early experimental measurements of bottom
production at the Fermilab Tevatron seemed to be significantly larger
than QCD calculations~\cite{Abe:1992fc,Acosta:2002qk,Abe:1992ww,Abe:1993sj,Abe:1995dv,Acosta:2001rz,Abachi:1994kj,Abbott:1999se,Abbott:1999wu}.  
It took
a while (see \cite{Cacciari:2004ur,Mangano:2004xr} for a review)
before these discrepancies could be resolved, through improvements in
the accuracy of both the experimental
measurements~\cite{Acosta:2004yw,Abulencia:2006ps,Aaltonen:2007zza} and the
theoretical predictions.  In particular, from the theoretical side,
the introduction of the ``Fixed Order + Next-to-Leading Log'' (\FONLL)
framework~\cite{Cacciari:1998it,Cacciari:2002pa} has shown that the
discrepancies are largely reduced if a consistent use is made of the
fragmentation function information coming from $e^+e^-$ data.  From
the experimental side, the extension of the $b$ production measurements
to very small transverse momenta~\cite{Acosta:2004yw} has demonstrated
good agreement with the fixed order QCD
calculation~\cite{Cacciari:2003uh} 
in a region where
theoretical uncertainties due to fragmentation effects have very
little relevance.

These successful comparisons notwithstanding, independent checks at a
larger centre-of-mass energy and with different measurements are of great
interest. Firstly, low-$p_T$ production at the LHC energies probes
values of the momentum fraction $x$ smaller than at the Tevatron, 
and challenges QCD in a
dynamical region where potentially large higher-order corrections need
to be resummed. This is particularly true of production at large
rapidity, which pushes $x$ of one of the two initial-state partons to
values of $x \lsim 10^{-4}$, and which can be studied by forward detectors
like LHCb and the ALICE muon spectrometer. Secondly, the greater beam
energy and the high luminosities of the LHC can push the kinematic reach to
much larger transverse momenta, exposing another interesting dynamical
regime, where the resummation of logarithms of big ratios (such as
$p_T^Q/m_Q$, $E_T^{jet}/m_Q$, $p_T^Q/p_T^{Q\bar{Q}}$, etc.) may become crucial.

In parallel with the most recent comparisons with Tevatron data mentioned 
above, new theoretical tools have become available. The successful
matching of next-to-leading order QCD calculations with parton shower
Monte Carlos (PSMCs) has led to the \MCNLO\
\cite{Frixione:2002ik,Frixione:2003ei} and the \PW\
\cite{Nason:2004rx,Frixione:2007nw} implementations, with matching to
\HW~\cite{Marchesini:1991ch,Corcella:2000bw} and to
\PY~\cite{Sjostrand:2006za}, which allow one to obtain predictions
for fully exclusive observables while guaranteeing that inclusive
 quantities retain full NLO accuracy. Predictions can now
therefore be evaluated within different frameworks (see also
\cite{Kniehl:2011bk, Kniehl:2012ti} for an independent analysis) and
compared among themselves as well as with the data.

The purpose of this paper is threefold. Firstly, we compare
theoretical predictions obtained within the \FONLL, \MCNLO\ and \PW\
frameworks for realistic observables (heavy mesons/hadrons, leptons
from heavy hadrons, $J/\psi$ from $b$-hadron decays). Secondly, we
compare these predictions, calculated within the cuts employed by the
experimental analyses, with the available data. Thirdly, we provide a
detailed public record for predictions that have been transmitted to
the experimental community over the past two years, and that have been
used in the comparisons with data that are documented in several
experimental papers.

\section{Description of the theoretical frameworks}

\subsection{FONLL}

In the \FONLL\ framework~\cite{Cacciari:1998it}, one
matches fixed next-to-leading order (NLO) QCD~\cite{Nason:1987xz,Nason:1989zy}
with all-order resummation to next-to-leading log (NLL) accuracy in the limit
where the transverse
momentum ($p_T$) of a heavy quark is much larger than its mass
($m$)~\cite{Cacciari:1993mq}. It allows one to calculate predictions for
one-particle inclusive distributions of a heavy quark (or heavy hadron), while
the degrees of freedom of the other particles in the event are
integrated over\footnote{We note that this implies that this approach
does not allow the study of correlations between the heavy quark 
and antiquark.}. The accuracy of the \FONLL\ calculation can be 
denoted as being NLO+NLL, with the  understanding that the logarithms resummed
up to next-to-leading accuracy are of the form $\alpha_s^n\log^n(p_T/m)$ and
$\alpha_s^n\log^{n-1}(p_T/m)$. The
perimeter of the meaning of the name ``\FONLL'' is often stretched to denote also
the determination, from $e^+e^-$ data~\cite{Cacciari:2005uk}, of the non-perturbative
fragmentation parameters necessary for a successful description of physical
differential distributions.

\FONLL\ has been used extensively to predict
bottom~\cite{Cacciari:2002pa, Cacciari:2003uh} and
charm~\cite{Cacciari:2003zu} 
production data at the Tevatron and at
RHIC~\cite{Cacciari:2005rk}. In all cases satisfactory agreement between theory
and data was found. The framework and the parameters employed in those
predictions have not been modified since (except for a more systematic
determination of the non-perturbative fragmentation parameters performed in
\cite{Cacciari:2005uk}, which confirmed earlier results).

A prediction for a single inclusive distribution (typically transverse momentum
($p_T$), rapidity ($y$) or pseudorapidity ($\eta$)) of a particle $\ell$
is obtained within \FONLL\ as a numerical convolution of a perturbative cross
section $d\sigma^{FONLL}_{Q}$ with a non-perturbative fragmentation function
$D^{NP}_{Q\to H_Q}$ and possibly a decay function $g^{weak}_{H_Q\to\ell}$
describing, for instance, the hadron weak decay into a lepton: 
\be
d\sigma_{\ell}^{FONLL} = d\sigma^{FONLL}_{Q} \otimes D^{NP}_{Q\to H_Q} \otimes 
g^{weak}_{H_Q\to\ell}\,.
\ee
The integral of the fragmentation functions $D^{NP}_{Q\to H_Q}$, for a given
heavy-flavoured hadron $H_Q$, will
be called the fragmentation fraction, $f(Q\to H_Q)$. 
The parameters for $D^{NP}_{Q\to H_Q}$ are best determined from $e^+e^-$ collisions data,
and a systematic survey is given in \cite{Cacciari:2005uk}. Weak decay spectra 
$g^{weak}_{H_Q\to \ell}$ and the corresponding branching ratios are similarly extracted from 
experimental data.

All the \FONLL\ predictions published in this paper, as well as others
corresponding to  different cuts, input parameters or distributions, can be
obtained from a publicly accessible web page~\cite{fonllform}.

\subsubsection{Non-perturbative fragmentation}
\label{sec:npfrag}

For completeness, we summarise here the parameters that we have used in this
paper.

For bottom production, the functional form chosen for the parametrization of
the non perturbative fragmentation function is a Kartvelishvili et al.
distribution~\cite{Kartvelishvili:1977pi}: \be
\label{eq:kart}
D^{NP}_{b\to H_b} = (\alpha + 1)(\alpha + 2) x^\alpha (1-x) \, .  \ee
We choose $m_b = 4.75$~GeV as central value for the bottom quark pole
mass, and the range $m_b = 4.5-5$~GeV for estimating the associated
uncertainty.  In this work we adopt the values of the fragmentation
parameter $\alpha$ obtained in ref.~\cite{Cacciari:2005uk}, namely
$\alpha = 24.2$ for $m_b = 4.75$~GeV, $\alpha = 26.7$ for $m_b =
4.5$~GeV and $\alpha = 22.2$ for $m_b = 5$~GeV.\footnote{These
  parameters differ from those employed in \cite{Cacciari:2002pa,
    Cacciari:2003uh, Cacciari:2005rk}, where $\alpha = 29.1$ was used
  as a central value with $m_b = 4.75$~GeV, together with $\alpha =
  34$ with $m_b = 4.5$~GeV and $\alpha = 25.6$ with $m_b = 5$~GeV
  (note that in the text of refs.~\cite{Cacciari:2003uh,
    Cacciari:2005rk} the values for $\alpha$ corresponding to $m_b =
  4.5$ and $m_b = 5$~GeV are unfortunately inverted, although the
  numerical results were obtained with the correct ones). The choice
  made in \cite{Cacciari:2002pa, Cacciari:2003uh, Cacciari:2005rk}
  corresponded to using a fit to the $N=2$ Mellin moment of the
  $e^+e^-$ data, whereas the values used in this paper correspond to a
  fit to the $N=5$ moment. The reason for a different choice is that
  the $N=5$ moment is more appropriate for describing steeply falling
  $p_T$ distributions like the ones found at large $p_T$ at the
  LHC. One should note, however, that switching from the old
  parameters set to the new one can be seen to lower the predictions
  by 5-10\% at most: it does not affect therefore the good agreement
  with the data found in the past.} These parameters were fitted in
\cite{Cacciari:2005uk} to the LEP data relevant to the production of a
mixture of $b$-hadrons~\cite{Heister:2001jg,Abbiendi:2002vt}.  No data
are available for the individual hadrons, for instance $B^{\pm}$ or
$B^0$. One is therefore forced to assume a similar non-perturbative
fragmentation for all of them. This assumption is likely fulfilled to
a large extent, with the possible partial exception of $\Lambda_b$
production, which however only contributes a minor fraction of the
$b$-hadron yield. We shall collectively denote the $b$-hadrons by
$H_b$, and assume that each $b$ quark eventually produces a
$b$-hadron, i.e. the fragmentation fraction $f(b\to H_b)$ is equal to
one (and equivalently for the $\bar b$). The function in
eq.~(\ref{eq:kart}) is normalised to one: when used for a specific
state the appropriate fragmentation fraction (e.g. $f(b\to B^+)$) must
additionally be provided.

For charm production the situation is more complex. On one hand, experimental
data are available for individual mesons ($D^*$, $D^\pm$, $D^0$ and $\bar D^0$, $D_s$).
On the other, at least partial theoretical understanding exists on the
differences and similarities in the fragmentation of a heavy quark into a
pseudoscalar ($D$) or a vector ($D^*$) mesons (see e.g. \cite{Braaten:1994bz}). 
This understanding can therefore
be exploited in order to minimize the number of parameters to be extracted from
experimental data. To this aim, in ref. \cite{Cacciari:2003zu} the
non-perturbative fragmentation functions into different charmed mesons were
constructed exclusively in terms of the $c\to D^*$ fragmentation, whose single
parameter was extracted from ALEPH $e^+e^-$ data~\cite{Barate:1999bg}. 
The decays of $D^*$ into $D$ states was
modeled theoretically, and the various branching ratios were extracted from
data. Primary $D$ production from $c$ fragmentation was described in terms of
the same non-perturbative parameter fitted to $c\to D^*$ data, though a
different functional form, as computed in \cite{Braaten:1994bz}, was used. The
detailed results for the non-perturbative fragmentation functions $D^{NP}_{c\to
D^*}$, $D^{NP}_{c\to D^+}$ and $D^{NP}_{c\to D^0}$ are given in eqs. (10), (9)
and (5) of \cite{Cacciari:2003zu} respectively, and we do not repeat them here
for brevity. They depend (besides the branching ratios of experimental origin)
on a single non-perturbative parameter $r$ determined from data. We
used the values $r = 0.1$ with $m_c = 1.5$~GeV, $r = 0.06$ with $m_c = 1.3$~GeV
and $r = 0.135$ with $m_c = 1.7$~GeV. These values also correspond to a $N=5$
fit as detailed in \cite{Cacciari:2005uk}.

\subsubsection{Theoretical uncertainties}
\label{sec:unc}

The `central' \FONLL\ prediction is computed by setting the
renormalisation and factorisation scales equal to the transverse mass,
$\mu_{R,F} = \mu_0 \equiv \sqrt{p_T^2 + m^2}$: here, $m$ and $p_T$ are 
the heavy {\em quark} mass and transverse momentum respectively. 
The theoretical uncertainty is estimated as a combination
of factorisation and renormalisation scale variations, heavy quark
mass variation, and uncertainty associated with Parton
Distribution Functions (PDFs). The three uncertainties will normally
be combined in quadrature.
\begin{itemize}
\item
In order to avoid accidental compensation between the $\mu_F$ and the $\mu_R$
dependence of the cross section, which may occur if the two scales are
set equal, we compute the scale uncertainty by varying them 
independently over the range $0.5 \le \xi_{R,F} \le 2$,
with the constraint $0.5 \le \xi_R/\xi_F \le 2$, where $\xi_{R,F} \equiv
\mu_{R,F}/\mu_0$. In practice, the cross section is evaluated (using the central
mass value and PDF set) in the seven points
\be
(\xi_{R},\xi_F) \;\in\; \{(1,1), (0.5,0.5), (2,2), (0.5,1), (1,0.5), (2,1),
(1,2)\}\, ,
\ee
and the envelope is taken. This envelope defines, at each point in the
distribution one is considering, the two extremes of
\be
d\sigma^{+\Delta_{+,scales}}_{-\Delta_{-,scales}}
\ee

\item
The three mass values $m=1.5$, 1.3 and 1.7 GeV are used for charm, and $m=4.75$,
4.5 and 5 GeV for bottom. Non-perturbative parameters are adjusted for each mass
to their appropriate value,\footnote{Note that this adjustment translates into a
non negligible reduction of the sensitivity of the hadron-level cross section on
the heavy quark mass value at large transverse momentum, as one may expect from
the fact that neither the heavy quark mass nor the non-perturbative
fragmentation are physical observables, and therefore their variations must
compensate each other in their interplay.} and the cross section is evaluated
setting $\xi_{R,F} = 1$ and with the central PDF set. The envelope is then taken,
defining 
\be
d\sigma^{+\Delta_{+,mass}}_{-\Delta_{-,mass}} \, .
\ee

\item
The PDF uncertainty, where available, is evaluated (setting $\xi_{R,F} = 1$, and
the mass equal to the central value) as suggested by the specific PDF set used (see e.g.
\cite{Nadolsky:2001yg}), 
yielding
\be
d\sigma^{+\Delta_{+,PDF}}_{-\Delta_{-,PDF}} \, .
\ee
Our default PDF set will be CTEQ6.6~\cite{Nadolsky:2008zw}, unless otherwise stated.
\end{itemize}

The full uncertainty band of the \FONLL\ predictions will be given by
\be
d\sigma^{+\Delta_{+}}_{-\Delta_{-}}
\ee
with
\be
\Delta_{\pm} = \sqrt{\Delta^2_{\pm,scales} + \Delta^2_{\pm,mass}+ \Delta^2_{\pm,PDF}}
\ee

\subsection{NLO+PS approaches}
Heavy flavour production has been available for quite some time in
NLO+PS (Next-to-leading order plus parton shower) models, namely in
\MCNLO~\cite{Frixione:2003ei} and in \PW~\cite{Frixione:2007nw}.
These programs, in conjunction with existing parton shower programs,
are capable of generating fully exclusive final states, maintaining the
next-to-leading order accuracy for inclusive observables. 

We remark that, unlike the \FONLL\ approach, these
generators resum only a subset of all contributions enhanced by large logarithms
of the ratio of the transverse momentum of the heavy quark over its
mass. In particular, the so called gluon splitting
and flavour excitation production mechanisms are implemented only
at order ${\cal O}(\alpha_S^3)$, i.e. the lowest order contributions to
these processes. In contrast, \FONLL\ includes for these contributions
all terms of order $\alpha_S^2 \times (\alpha \log p_T/m)^n$ and
$\alpha_S^3 \times (\alpha_S \log p_T/m)^n$. On the other hand, at small
and moderate transverse momenta, the NLO+PS approaches are superior to
\FONLL, since they have the same accuracy in this region but, unlike \FONLL,
provide a complete and fully exclusive description of the final state.

In the present comparison, the NLO+PS methods should be viewed as
approaches that work at small and moderate energies, and that are
bound to fail at some large transverse momentum scale. To some extent,
the present work helps to assess the range of validity of these
approaches.

It should be also kept in mind that the non-perturbative part of the
shower that leads to the formation of the $b$ hadrons is handled by
the Parton Shower program alone. The corresponding parameters are
tuned using final-state observables reconstructed with particles
emerging from the parton shower, but with the hard production cross
section of the shower Monte Carlo, i.e. not that of the
next-to-leading order ones used in the NLO+PS methods. This is in
contrast with what is done in \FONLL, where use is made of a
non-perturbative fragmentation function which is fitted to $e^+e^-$
data with a theoretical calculation based on the very same underlying
\FONLL\ approach.

We point out that the parameters that control cluster decays in
Fortran \HW~\cite{Corcella:2002jc} ({\tt PSPLT}, {\tt CLDIR}, and {\tt
  CLSMR}) may be assigned (depending on the tuning adopted) specific
values for $b$-flavoured clusters. On the other hand, $c$-flavoured
cluster decays are treated in the same manner as light quark
ones. Although one can envisage to relax this constraint, and to
introduce and tune parameters relevant to $c$-flavoured clusters only,
such a possibility is not given in the official \HW\ versions, and has
not been considered here.

\subsubsection{MC@NLO}
\label{sec:mcatnlo}

The \MCNLO\ formalism has been introduced in ref.~\cite{Frixione:2002ik},
and aims at a consistent matching between NLO QCD corrections for a
given process, and parton showers.
The relevant technical details are given in the quoted references,
and we shall omit them here.
We limit ourselves to recall that in the context of \MCNLO\ the matching 
prescription amounts to modifying the short-distance cross sections 
relevant to the NLO computation, by including the so-called Monte Carlo 
(MC) subtraction terms, that are responsible for removing any double 
counting at the NLO. The MC subtraction terms can be computed in a 
process-independent manner, but they are still dependent on the particular 
PSMC 
one adopts for the shower phase. In other words, each PSMC requires 
a set of MC subtraction terms, which can 
be obtained by formally expanding the PSMC results to the same order
in $\as$ as the corresponding NLO contribution to the parton-level 
cross section (i.e. that of the real-emission matrix elements).
Furthermore, their structures are such that all non trivial process-specific 
information is contained in the Born matrix elements. These matrix elements 
are multiplied by kernels whose analytic forms depend solely on the 
shower variables used by the PSMC to generate the elementary branchings,
and on the identities of the partons involved in such branchings.

Although ref.~\cite{Frixione:2002ik} formulated the solution of
NLO+PS matching in general terms, practical applications there
and in subsequent papers have been restricted to the choice of
Fortran \HW~\cite{Marchesini:1992ch,Corcella:2001bw,Corcella:2002jc}
as PSMC. Recently, MC subtraction terms have been computed which are
relevant to the matching with \HWpp~\cite{Bahr:2008pv} and, for 
processes that feature only initial-state emissions, with
\PY\ 6.4~\cite{Sjostrand:2006za} (see refs.~\cite{Frixione:2010ra}
and~\cite{Torrielli:2010aw} respectively\footnote{The matching 
with \PY\ (including the $p_{\sss T}$-ordered shower versions) for
a generic process is being achieved in the context of a\MCNLO\ --
see {\tt http://amcatnlo.cern.ch} for more information on this project.}).
In the present work, the \MCNLO\ results have been obtained with
Fortran \HW\ (v6.520, which, for the processes considered here,
gives results identical to those of v6.510), 
using the implementation of $Q\bar{Q}$ production
presented in ref.~\cite{Frixione:2003ei}. We point out that, although
such an implementation technically only requires  the mass of
the heavy quark to be larger than $\Lambda_{\rm QCD}$, the conservative
choice has been made of not including $c\bar{c}$ production in
the public \MCNLO\ package. This is due to the fact that, because
of the relative lightness of the charm quark, contributions may be
important where a light-parton scattering such as $gg\to gg$ initiates
showers that eventually feature a $g\to c\bar{c}$ branching. These
kind of contributions are not included in an NLO+PS matched computation,
unless the aforementioned branching is the first occurring in a given
shower. Although this mechanism is relevant also in the case of 
$b\bar{b}$ production, it is expected to be less important there
than in the case of $c\bar{c}$ final states. It is therefore interesting
to compare NLO+PS predictions with \FONLL, which does take correctly
into account the contributions discussed above. The default scale
choice for $Q\bar{Q}$ production in \MCNLO\ is
\beq
\mu_0^2=\frac{1}{2}\left(m_{\sss T}^2(Q)+m_{\sss T}^2(\bar{Q})\right)\,,
\phantom{aaaa}
m_{\sss T}^2=p_{\sss T}^2+m^2\,,
\eeq
where $p_{\sss T}$ is the transverse momentum w.r.t.~the beam line
(note that, in the real-emission contributions to the NLO cross section,
the transverse momentum of the $Q$ is generally not equal to that of
the $\bar{Q}$).

\subsubsection{POWHEG}
\label{sec:powheg}

The \PW\ implementation of heavy flavour production is described in
detail in ref.~\cite{Frixione:2007nw}.  It is now available at the
{\tt \PW\ BOX} website, {\tt http://powhegbox.mib.infn.it/}. 
It can be used to generate events with
either $t\bar{t}$, $b\bar{b}$ or $c\bar{c}$ pairs, and it is based
upon the heavy flavour production next-to-leading order calculation of
refs.~\cite{Mangano:1991jk,Nason:1989zy,Nason:1987xz}.  The output is
an event file (in the Les Houches format~\cite{Alwall:2006yp}), that
can be fed through any Shower Monte Carlo program that complies with
the requirements of the Les Houches Interface for User Processes (LHIUP),
like the Fortran and C++ versions of \PY\ and \HW,
in order to generate complete events.

The default scale choice used in this implementation is given by
$\sqrt{p_T^2+m^2}$, where $p_T$ is the transverse momentum of the
heavy flavour in the underlying Born configuration. This is the
structure of the $Q\bar{Q}$ event before the hardest radiation has
been generated, as defined in
refs. \cite{Alioli:2010xd,Frixione:2007vw}.  It is obtained as
follows. One performs a longitudinal boost of the heavy quark pair
system such that the system has zero rapidity after the boost, then a
transverse boost is performed such that the transverse momentum of the
system vanishes, and then the inverse of the initial longitudinal
boost is performed. Thus, although the total cross section in \PW\
corresponds to the NLO result, as in \MCNLO, one cannot expect the two
generators to yield identical cross sections because of the different
scale choice.

\begin{figure}[t]
\begin{center}
\includegraphics[height=0.49\textwidth,angle=-90]{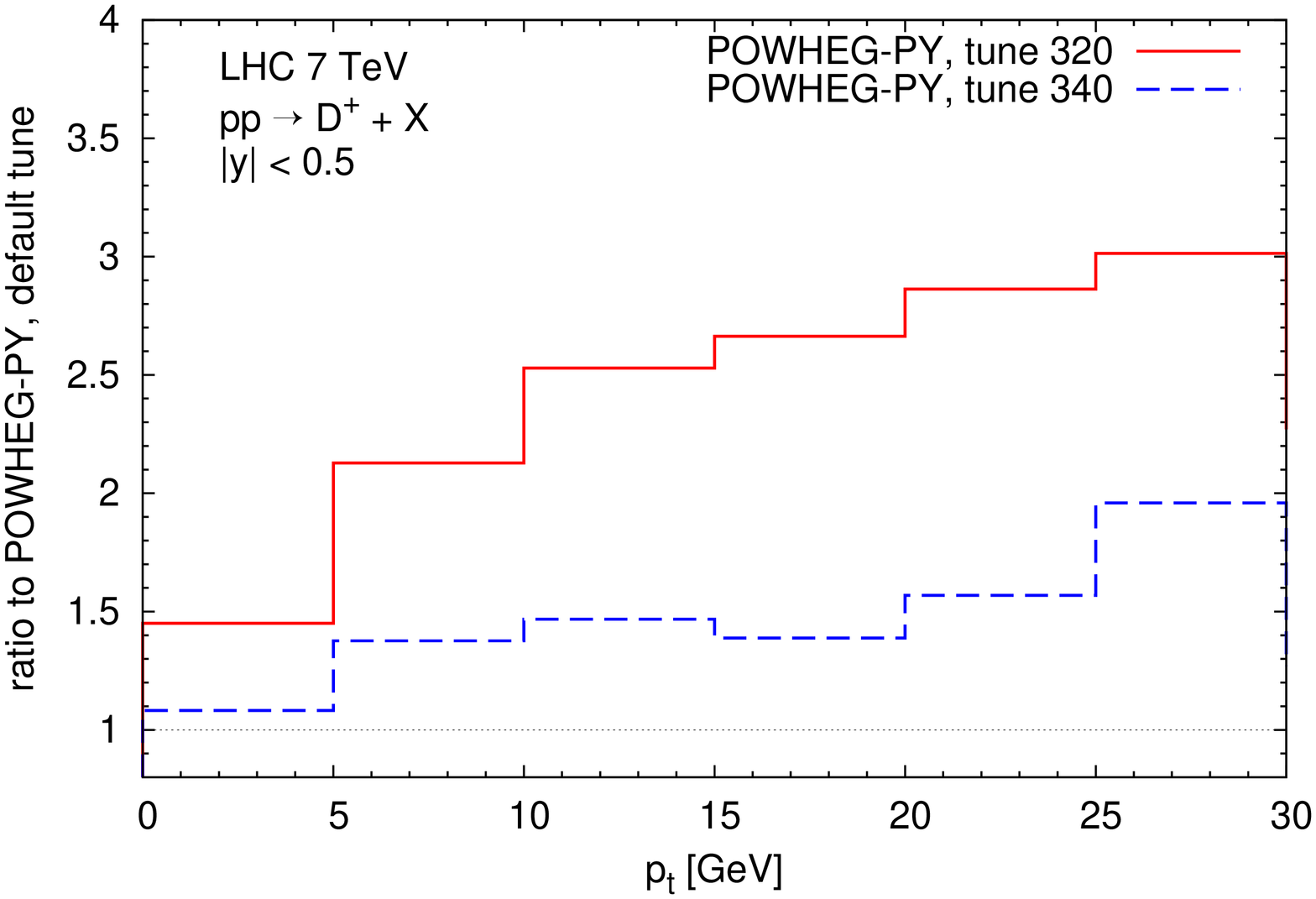}
\includegraphics[height=0.49\textwidth,angle=-90]{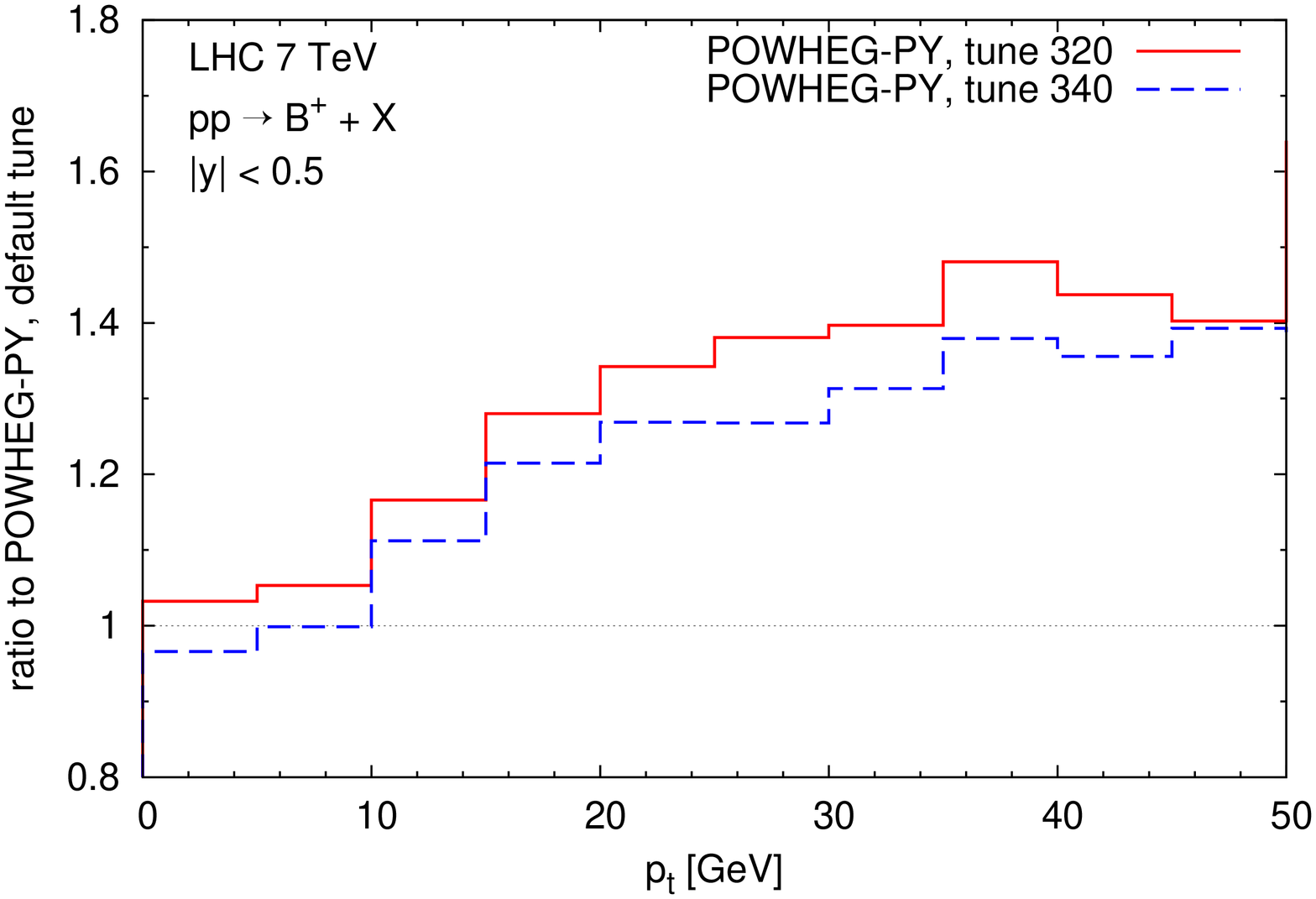}
\caption{\label{fig:tunes} Sensitivity of \PW\ with \PY\ predictions to the tune
used in \PY\, for $D^+$ (left plot) and $B^+$ (right plot) transverse momentum distributions in
the central rapidity region. At large rapidities, we observe effects of similar
  size or larger.}
\end{center}
\end{figure}

\PW\ has been interfaced to \PY\ version 6.4.25,
and \HW\ 6.510, keeping always the default values of the parameters
in the Monte Carlo. A non negligible sensitivity to the Monte Carlo tune
is expected for the observables considered in this work. An example is given in
figure~\ref{fig:tunes}, where $D^+$ and $B^+$ production 
as predicted by \PW\ with \PY\ is studied for different \PY\ tunes in $|y|<0.5$. At large rapidities
the sensitivity to the \PY\ tune is similar, if not larger.
A complete study of the tune dependence is outside the scope of the present
paper, and all Monte Carlo results presented in this paper will employ the default
tune, unless explicitly stated.

\section{Numerical predictions and comparisons}

\subsection{Open charm production}

\begin{figure}[t]
\begin{center}
\includegraphics[height=0.48\textwidth,angle=-90]{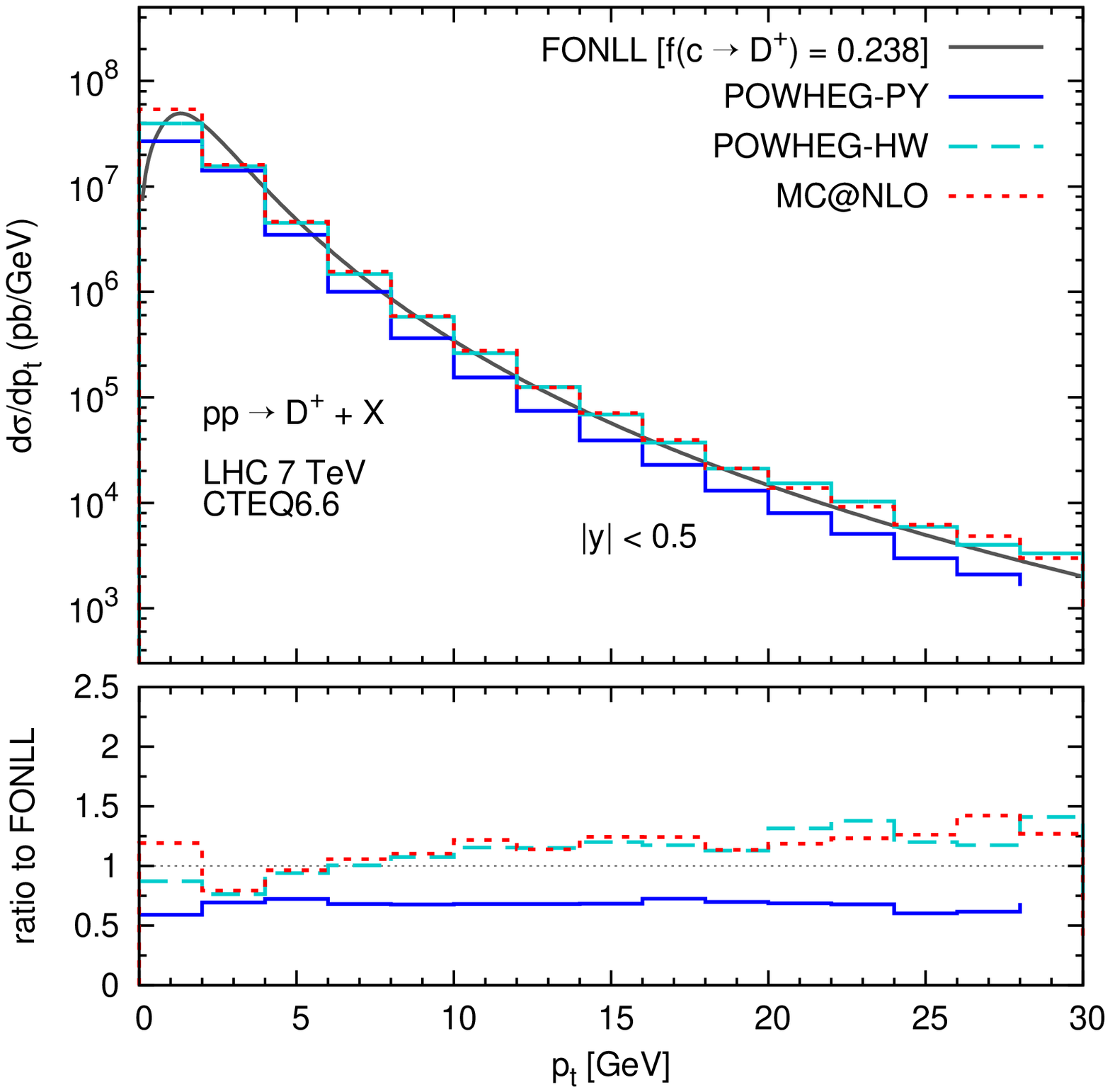}
\hfill
\includegraphics[height=0.48\textwidth,angle=-90]{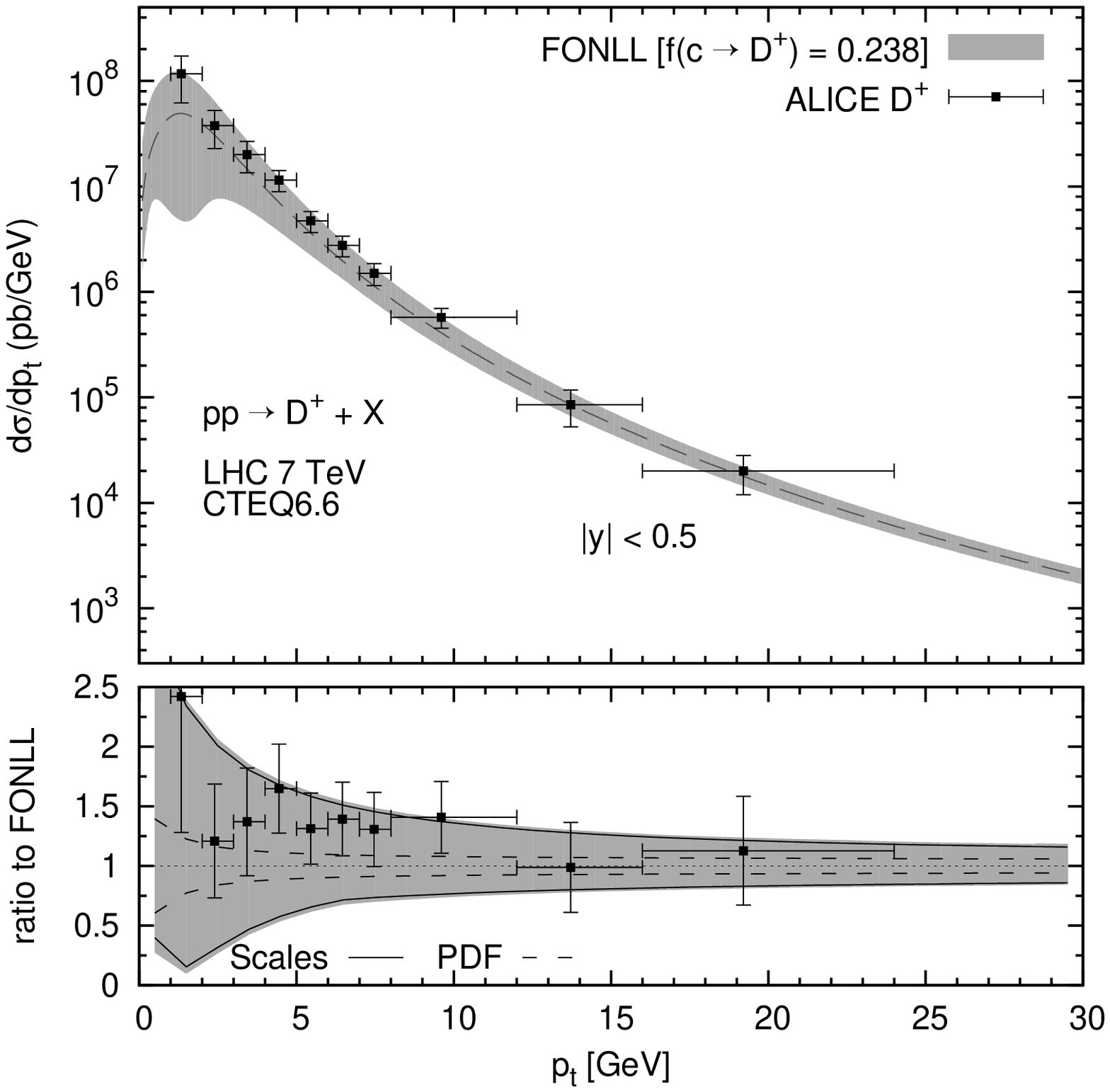}
\caption{\label{fig:dplus} Transverse momentum distribution of
  $D^+$ mesons at central rapidity, $|y|<0.5$. Left plot: comparison
  among the central predictions of our four benchmark calculations,
  \FONLL, \MCNLO, and \PW\ with \PY\ or \HW\ showers. Right plot:
  theoretical systematics for the \FONLL\ calculation, and the
  comparison with data from
  ALICE~\cite{ALICE:2011aa}. For the systematics we show the
  individual scale and PDF components, as well as the combined
  total (which includes mass variation, as described in the text).}
\end{center}
\end{figure}

\begin{figure}[t]
\begin{center}
\includegraphics[height=0.48\textwidth,angle=-90]{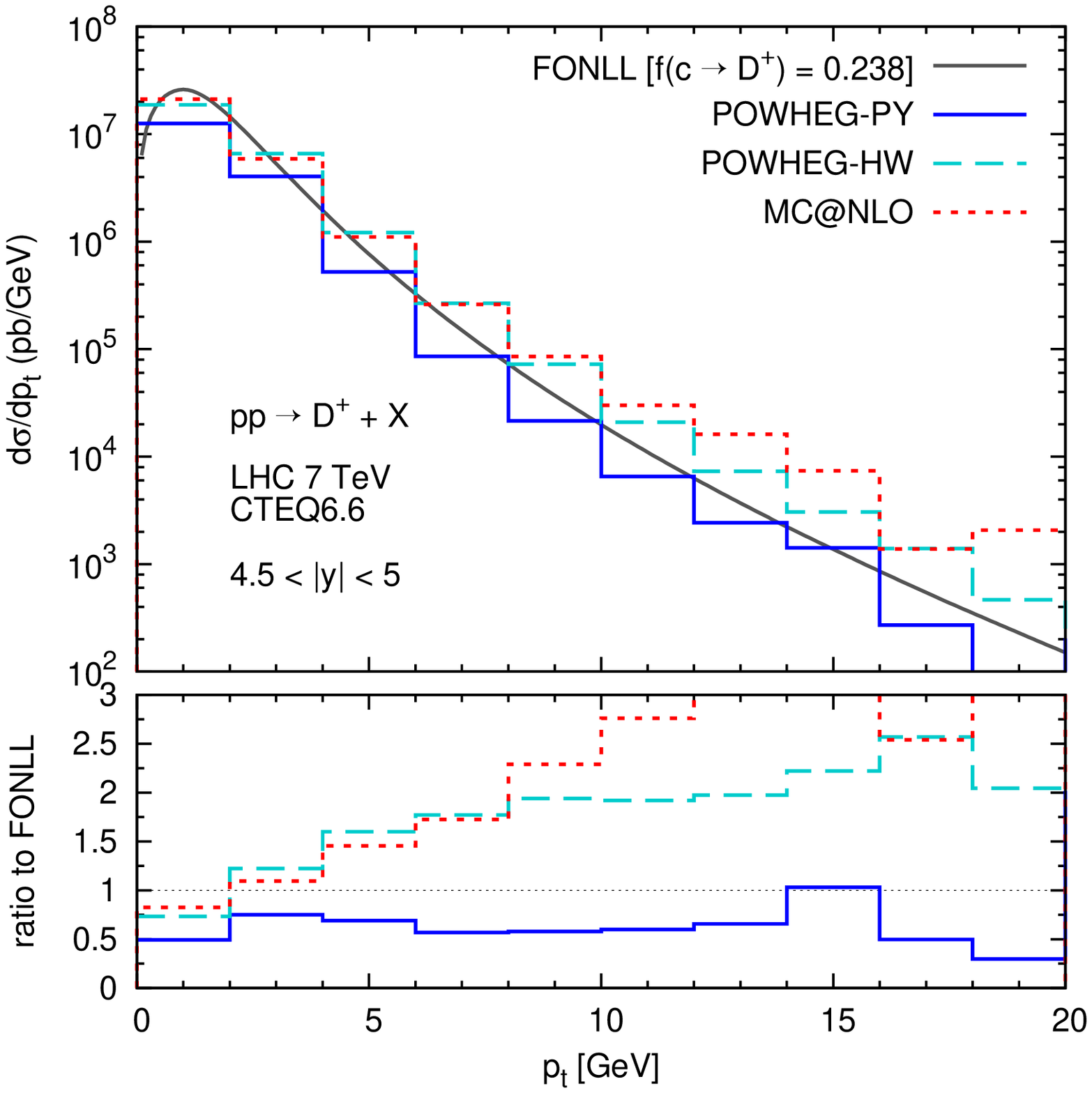}
\hfill
\includegraphics[height=0.48\textwidth,angle=-90]{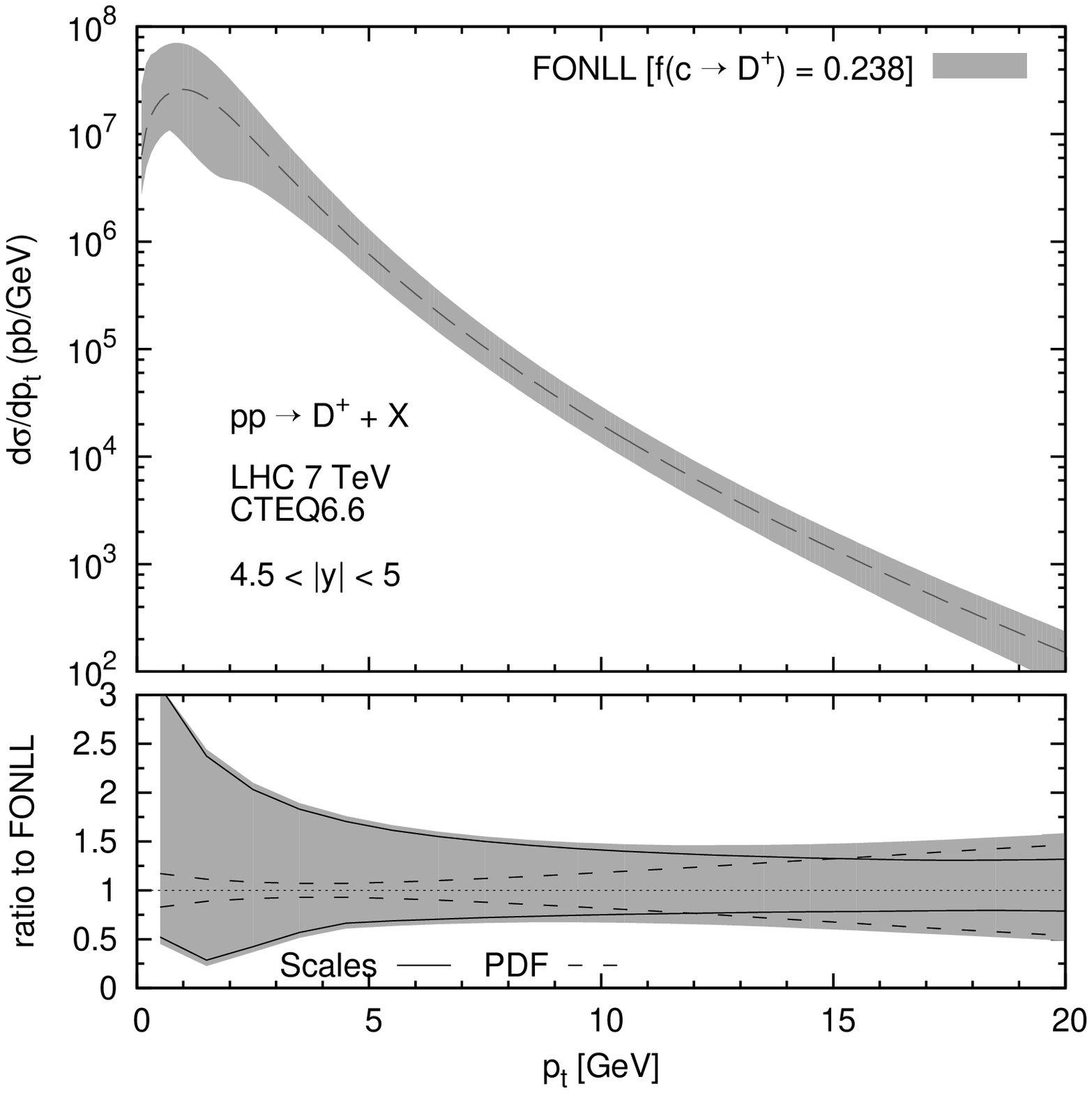}
\caption{\label{fig:dplus_fwd} Same theoretical distributions as in
  Fig.~\ref{fig:dplus}, for the forward rapidity region $4.5<|y|<
  5$. }
\end{center}
\end{figure}

We begin by considering the production of $D$ mesons. The \FONLL\ framework
and its non-perturbative parameters are described in
Section~\ref{sec:npfrag}. In the following we concentrate on the production of $D^+$
mesons (the others being largely similar) and we use a value of the
fragmentation fraction of $f(c\to D^+)=0.238$ with \FONLL. 
The left plot of Fig.~\ref{fig:dplus}
shows the central \FONLL\ prediction, compared with \MCNLO, \PW\ with \PY\
(denoted by \PWPY) and \PW\ with \HW\ (denoted by
\PWHW). The right plot shows the scale and PDF contributions to
the \FONLL\ uncertainty, as well as the overall systematics. The ALICE
data from ref.~\cite{ALICE:2011aa} are also compared to the
theoretical prediction, showing a good agreement within the large
systematics. We note that all four theoretical predictions are in
fair agreement with each other within the overall uncertainty. The
large dependence, in shape as well as overall normalisation, of the Monte Carlo 
results on the specific tune, as shown in section~\ref{sec:powheg}, should also 
be kept in mind. 
Differences appear, on the other hand, 
in the forward region $4.5<|y|<5$, as shown in
Fig.~\ref{fig:dplus_fwd}. \PWPY\ agrees well with \FONLL, while the
shower evolution carried out with \HW\ (whether in the \MCNLO\ or in
the \PW\ framework) leads to a harder spectrum. These differences largely exceed the
systematics quoted from the \FONLL\ calculation, in particular if we
consider that the sightly bigger uncertainty at large $p_T$ is mostly
due to the PDFs, and is therefore entirely correlated among the
different predictions.
It will be very interesting to see the first data on forward $D^+$
production at large $p_T$ from LHCb.

Additional data on $D^0$ and $D^{*+}$ production in the central
region, and comparisons with the \FONLL\ predictions,
are reported by the ALICE Collaboration in Fig.~5 of
\cite{ALICE:2011aa}. Similar data, at $\sqrt{S}=2.76$~TeV, are
reported in~\cite{ALICE:2012sx}. 
Preliminary data from ATLAS are also available,
and compared to \FONLL\ in Fig.~1 and 2 of~\cite{ATLAS-Dmesons}. In either
case a fair agreement is found, with the data mostly centred on the upper
edge of the theoretical uncertainty band, mirroring quite closely the
comparisons with the Tevatron $D$ mesons data~\cite{Cacciari:2003zu}.
This may indicate a preference of the data for a value of the charm mass
smaller than our default of 1.5~GeV.

\subsection{Open bottom production from inclusive and
  fully-reconstructed $H_b\to D+X$ decays}

\begin{figure}[t]
\begin{center}
\includegraphics[height=0.48\textwidth,angle=-90]{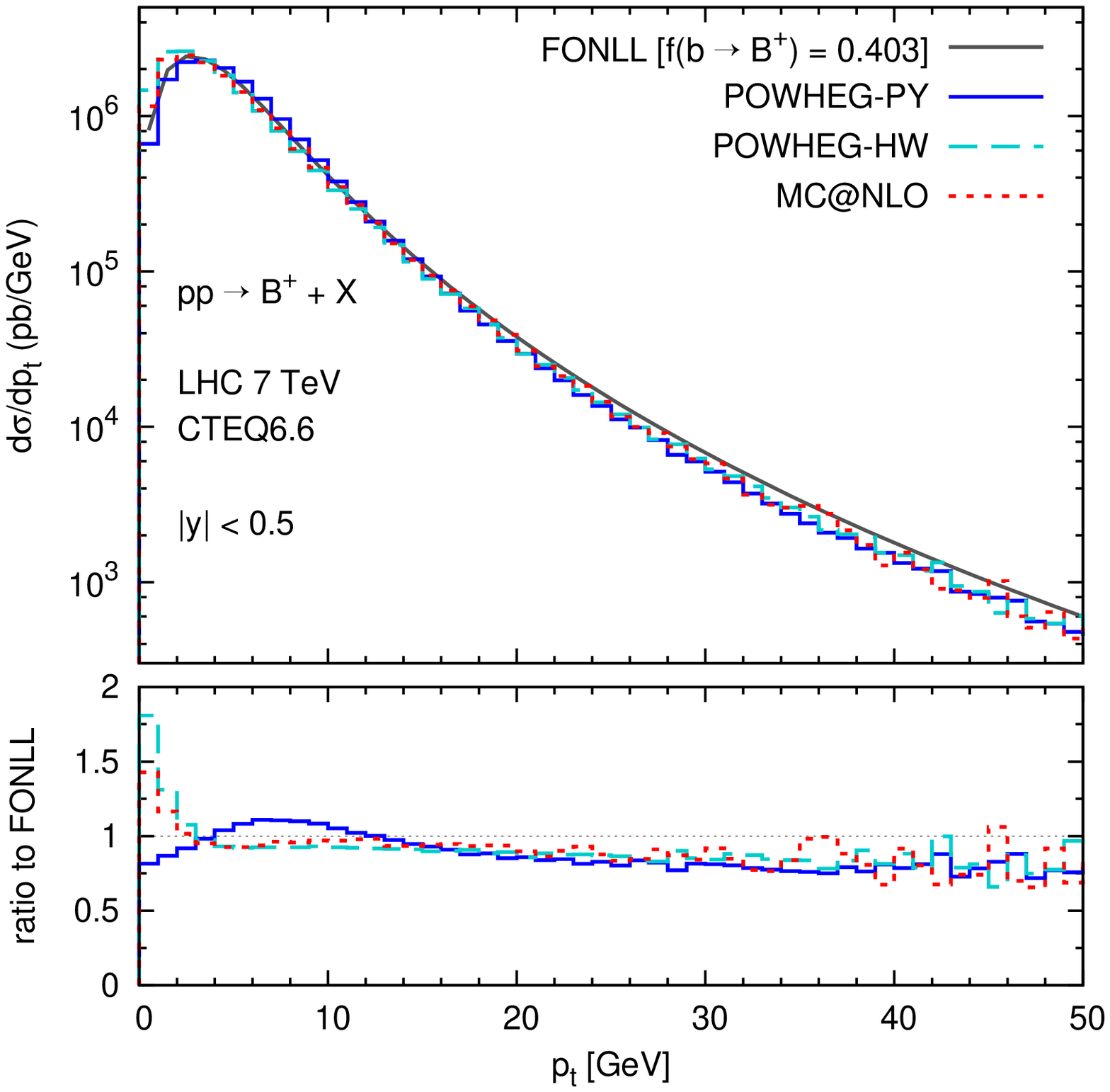}
\hfill
\includegraphics[height=0.48\textwidth,angle=-90]{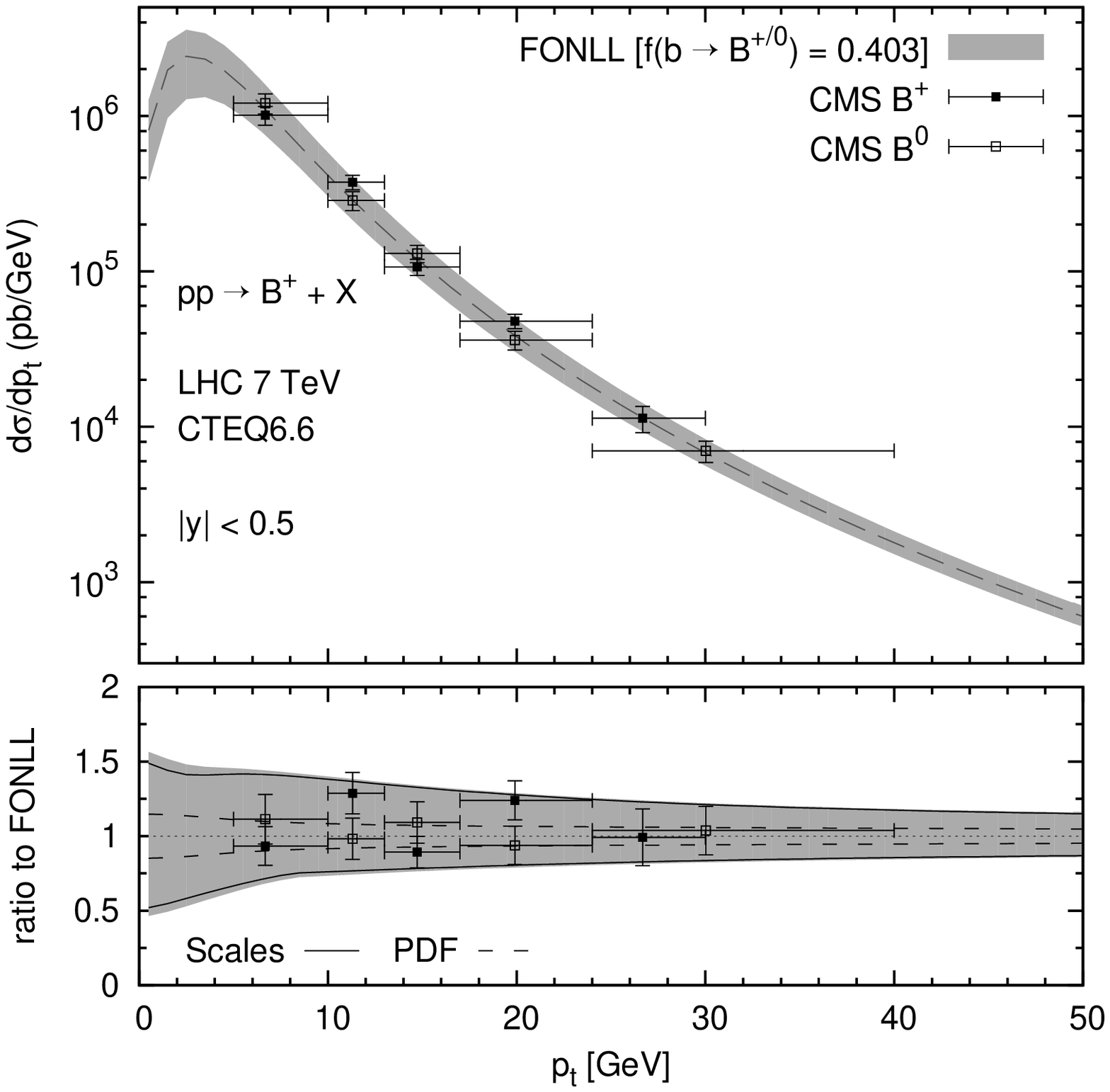}
\caption{\label{fig:bplus} 
Transverse momentum distribution of
  $B^+$ mesons at central rapidity, $|y|<0.5$. Left plot: comparison
  among the central predictions of our four benchmark calculations,
  \FONLL, \MCNLO, and \PW\ with \PY\ or \HW\ showers. Right plot:
  theoretical systematics for the \FONLL\ calculation, and the
  comparison with data from
CMS~\protect\cite{Khachatryan:2011mk,Chatrchyan:2011pw}, rescaled to the $|y^B|<0.5$
region.
For the systematics we show the
  individual scale and PDF components, as well as the combined
  total.}
\end{center}
\end{figure}

\begin{figure}[t]
\begin{center}
\includegraphics[height=0.48\textwidth,angle=-90]{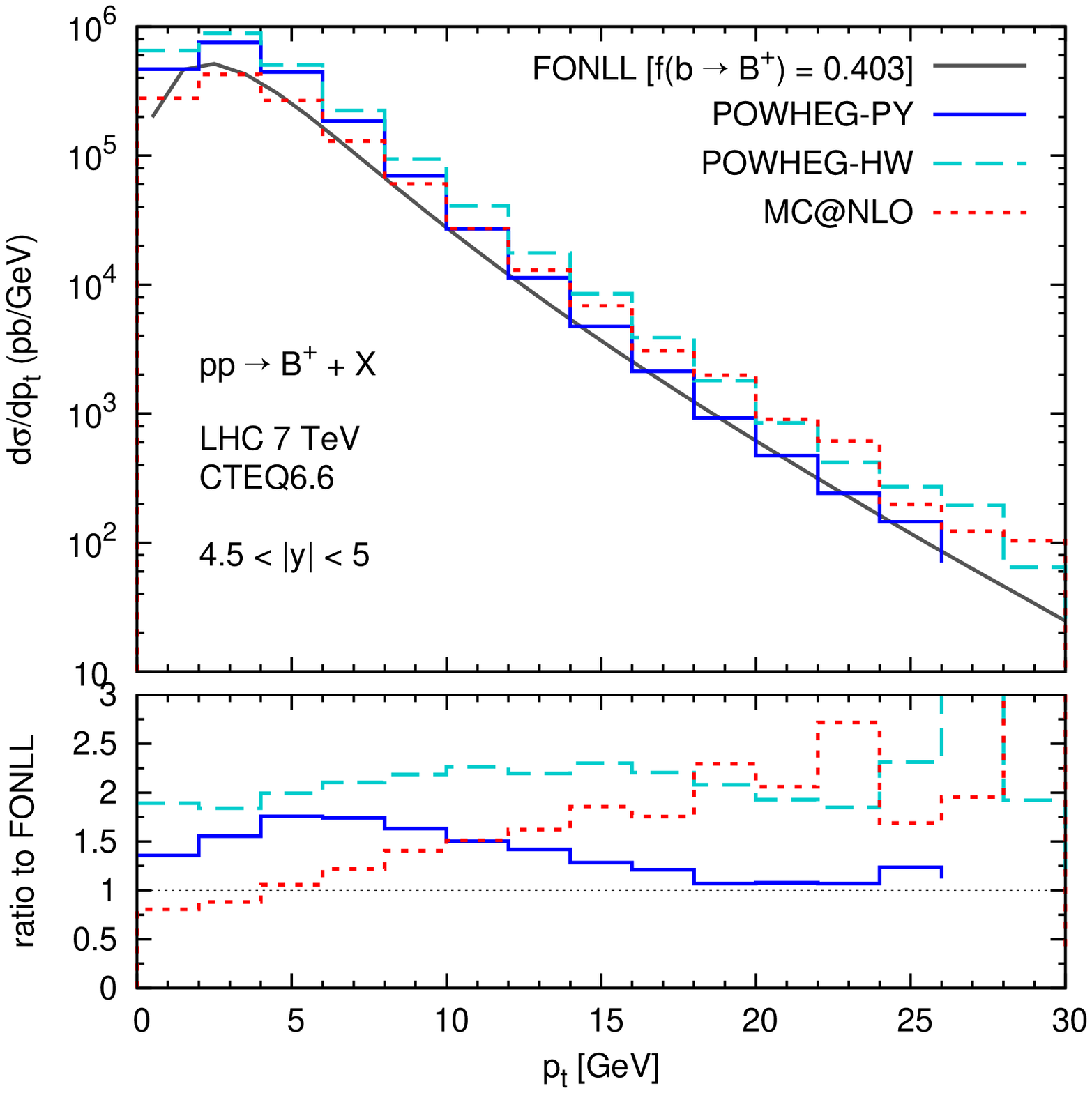}
\hfill
\includegraphics[height=0.48\textwidth,angle=-90]{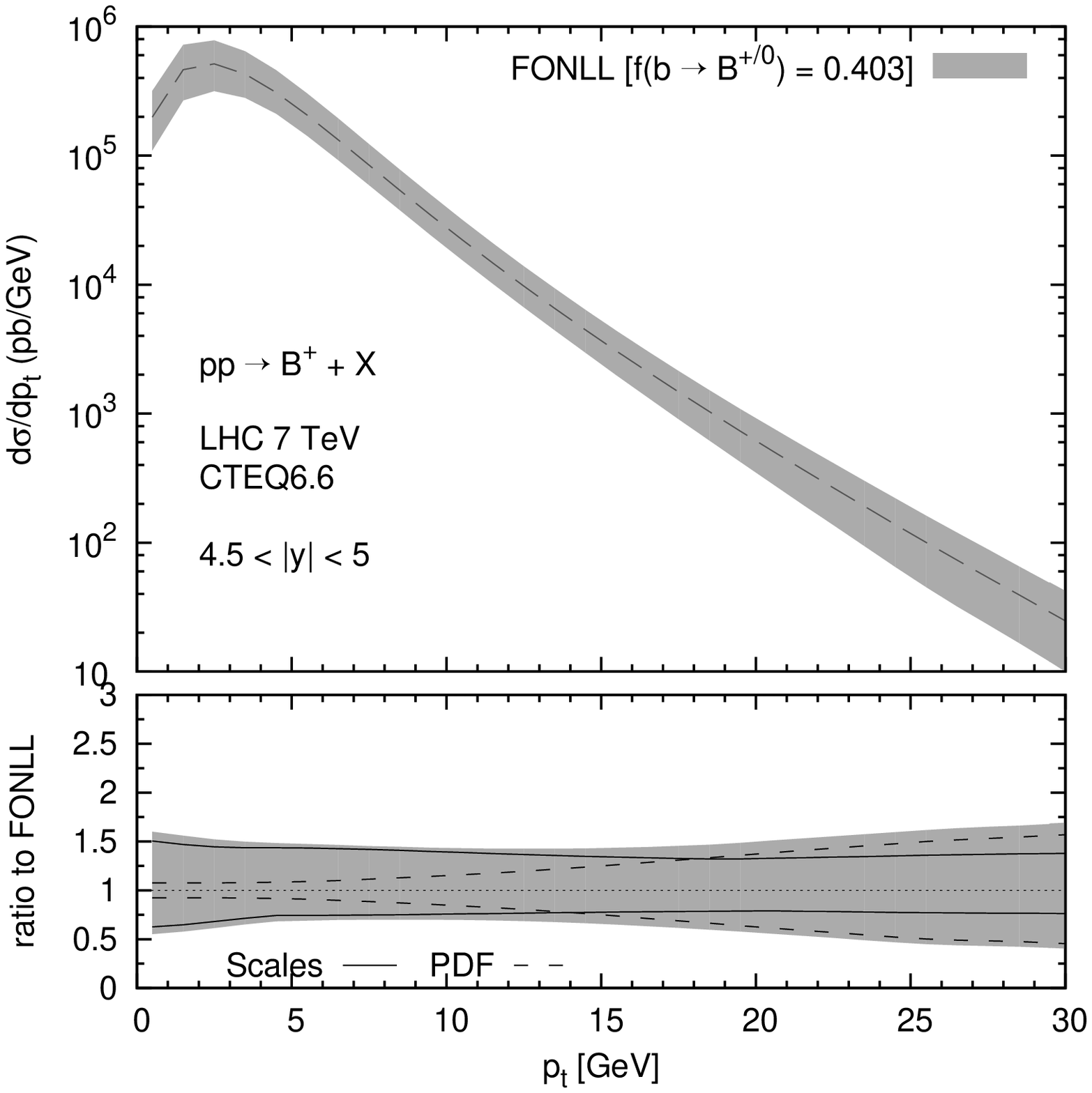}
\caption{\label{fig:bplus_fwd} Same theoretical distributions as in
  Fig.~\ref{fig:bplus}, for the forward rapidity region $4.5<|y|< 5$.}
\end{center}
\end{figure}

The theoretical predictions for central production of $B^+$ mesons
($\vert y \vert <0.5$) are presented in
Fig.~\ref{fig:bplus}. The quality of the agreement among the various
predictions is similar to the one seen in the $D^+$ case
above. Other features, like -- at large rapidity -- the progressively larger theoretical
uncertainty as a consequence of a larger PDF
uncertainty, or the larger variance between the NLO+PS and the \FONLL\ predictions
(see Fig.~\ref{fig:bplus_fwd}), are also similar. The potential
sensitivity of the Monte Carlo results to the specific tune used (though smaller
than in the $D$-mesons case, see figure~\ref{fig:tunes}), should be kept in mind in
this case too.

The first measurement of $b$-hadron production at the LHC was performed
by the LHCb Collaboration \cite{Aaij:2010gn}. The pseudorapidity
distribution in the region $2\le \eta\le 6$ was shown to be in good
agreement with NLO and \FONLL\ predictions (see Fig. 5 of
\cite{Aaij:2010gn}).  The measured total cross section in this region
was found to be (averaging over $b$ and $\bar b$
hadrons)\footnote{Note that, as explained in \cite{Aaij:2010gn}, the
  measurement can change to $89.6 \pm 6.4\pm 15.5~\mu b$ if $b$-hadron
  fractions measured at the Tevatron rather than those measured at LEP
  are used in converting the number of events to an $H_b$ cross
  section.}
\be
\sigma^\mathrm{LHCb}(pp\to H_b, 2\le \eta\le 6) = 75.3 \pm 5.4\pm 10.0~\mu b \\
\ee
to be compared with the \FONLL\ prediction\footnote{Here and elsewhere,
  where the quoted \FONLL\ predictions already appeared in experimental
  papers. They were originally calculated according to the
  framework defined in this paper, and provided as private
  communications to the experimental
  collaborations.}
\be
\sigma^\mathrm{FONLL}(pp\to H_b, 2\le \eta\le 6) = 
70.8~^{+33.3}_{-24.4}~\mu b\,. \\
\ee
As remarked above, the contribution of $b$-baryons are included in
this theoretical value 
assuming they fragment like the $B$ mesons.

More recently, LHCb has published 
the measurement of the $B^\pm$ production cross section in the $2 < y < 4.5$ 
rapidity region~\cite{Aaij:2012jd}, using fully reconstructed $B^\pm
\to J\!/\!\psi K^\pm$ decays:
\be
\sigma^\mathrm{LHCb}(pp\to B^\pm, 0 < p_T < 40~\mathrm{GeV}, 2 < y < 4.5) = 
41.4 \pm 1.5\pm 3.1~\mu b 
\ee
to be compared with the \FONLL\ prediction 
(which includes a fragmentation fraction $f(b\to B^{-})$ = 0.403)
\be
\sigma^\mathrm{FONLL}(pp\to B^\pm, 0 < p_T < 40~\mathrm{GeV}, 2\le y \le 4.5) = 
40.1~^{+19.0}_{-14.5}~\mu b\,.
\ee
Good agreement in seen also in the differential $p_T^{B^\pm}$ spectra,
in the range up to 40~GeV, as shown in Fig.~2 of~\cite{Aaij:2012jd}.

The CMS collaboration has published measurements of transverse
momentum and rapidity distributions of
$B^+$~\cite{Khachatryan:2011mk}, $B^0$~\cite{Chatrchyan:2011pw} and
$B^0_s$~\cite{Chatrchyan:2011vh} mesons. They report, for the
total visible cross sections at 7 TeV LHC,
\ba
&&\!\!\!\!\!\!\sigma^\mathrm{CMS}(pp\to B^0,~p_T^B > 5~\mathrm{GeV},~|y^B| < 2.2) = 33.2 \pm 2.5\pm 3.5~\mu b \\
&&\!\!\!\!\!\!\sigma^\mathrm{CMS}(pp\to B^+,~p_T^B > 5~\mathrm{GeV},~|y^B| < 2.4) = 28.1 \pm 2.4\pm 2.0 \pm 3.1~\mu b \\
&&\!\!\!\!\!\!\sigma^\mathrm{CMS}(pp\to B^0_s,~8 < p_T^B < 50~\mathrm{GeV},~|y^B| < 2.4)\times \mathrm{BR}(B^0_s\to J\!/\!\psi\,\phi) = 6.9 \pm 0.6\pm 0.6~nb\,,
\ea
to be compared with the \FONLL\ predictions
\ba
&&\!\!\!\!\!\!\sigma^\mathrm{FONLL}(pp\to B^0,~p_T^B > 5~\mathrm{GeV},~|y^B| <
2.2) = 25.5~^{+10.5}_{-7.1}~\mu b 
\label{CMSfonll1}\\
&&\!\!\!\!\!\!\sigma^\mathrm{FONLL}(pp\to B^+,~p_T^B > 5~\mathrm{GeV},~|y^B| <
2.4) = 27.2~^{+11.2}_{-7.5}~\mu b 
\label{CMSfonll2}\\
&&\!\!\!\!\!\!\sigma^\mathrm{FONLL}(pp\to B^0_s,~8 < p_T^B < 50~\mathrm{GeV},~|y^B| < 2.4)\times \mathrm{BR}(B^0_s\to J\!/\!\psi\,\phi) = 4.5~^{+1.7}_{-1.1} \pm
1.6~nb\,.~
\label{CMSfonll3}
\ea
The \FONLL\ uncertainties in eqs.~(\ref{CMSfonll1})--(\ref{CMSfonll3}) are 
due to renormalisation and factorisation scales, heavy quark masses and PDF, as
detailed in section~\ref{sec:unc}. Of these three sources, the first is 
largely dominant. 
In the \FONLL\ predictions the fragmentation fractions $f(b\to B^{0})=f(b\to B^{-})$ = 0.403 and 
$f(b\to B^{0}_s)$ = 0.11~\cite{Nakamura:2010zzi} have been included. 
These values, leading to a ratio $f_s/f_d=0.273$, are consistent with
the recent direct measurement of this ratio by
LHCb~\cite{Aaij:2011jp}: $f_s/f_d=0.267{+0.021 \atop -0.020}$. Additionally, 
BR$(B^0_s\to J\!/\!\psi\,\phi) = (1.4 \pm 0.5)\times
10^{-3}$~\cite{Nakamura:2010zzi}  has been used, and the second uncertainty 
in the \FONLL\ prediction of eq.~(\ref{CMSfonll3}) for $B^0_s$ production
reflects the large uncertainty of this measured branching ratio.

A good agreement with the experimental measurements within the respective uncertainties can
be observed, confirming the latest comparisons between theory and Tevatron data which showed
no significant excess in bottom hadroproduction compared to theoretical predictions.
The $p_T$ spectra measured by CMS are compared to various predictions
in Fig.~2 of ref.~\cite{Chatrchyan:2011pw} and Fig.~2 of
ref.~\cite{Khachatryan:2011mk}. The comparison of these data, rescaled to the rapidity region
$\vert y \vert < 0.5$, with the \FONLL\ predictions is shown in the
right panel of Fig.~\ref{fig:bplus}. 

\subsection{Open bottom production from inclusive $H_b\to J\!/\!\psi$ and $H_b\to \psi(2S)$ decays}

The predictions for non-prompt $J\!/\!\psi$ and $\psi(2S)$,
i.e. coming from $b$-hadron decays, is obtained in \FONLL\ by
convoluting the distribution for $b$-hadron production with a
phenomenological spectrum, obtained from experimental 
data~\cite{Aubert:2002hc} (that describes
the momentum distribution of the $J\!/\!\psi$ and the $\psi(2S)$ in
$B$-meson decays), and by multiplying it by the appropriate branching fraction
(again obtained from experimental measurements).  In using decay
spectra from $B$ meson decays and applying them to all $b$-hadron
decays we make the reasonable assumption that a potential difference in
the small fraction ($\sim$10-20\%) of non-meson $b$-hadrons is going
to be inconsequential. Moreover, it should be noted that strictly
speaking ref.~\cite{Aubert:2002hc} gives the momentum of the
quarkonium in the $\Upsilon(4S)$ rest frame rather than in the
$B$-meson one. The difference has a small rms spread (0.12 GeV
according to \cite{Aubert:2002hc}) and is not expected to affect
significantly the final quarkonium spectrum in our convolution. In
the following we shall use the branching ratio BR$(b\to h)$, defined
as BR$(b\to h)=\sum_{H_b} \, f(b\to H_b) \times \mathrm{BR}(H_b \to
h+X)$, summed over all relevant b-hadrons $H_b$.

The fit to the $B\to J\!/\!\psi$ decay spectrum from Fig. 6
of~\cite{Aubert:2002hc} had been performed in preparation of
ref.~\cite{Cacciari:2003uh}, where non-prompt  $J\!/\!\psi$ distribution in $p\bar
p$ collisions at the Fermilab Tevatron had been calculated, and found in good
agreement with the data, as well as with the \MCNLO\ predictions (see Fig. 5 of
\cite{Cacciari:2003uh}). This paper uses the same spectrum. The fit to the 
$\psi(2S)$ spectrum from Fig. 8 of~\cite{Aubert:2002hc} has instead been performed
for this paper, and  predictions for the non-prompt $\psi(2S)$ within \FONLL\ are
given here for the first time.

\begin{figure}[t]
\begin{center}
\includegraphics[height=0.48\textwidth,angle=-90]{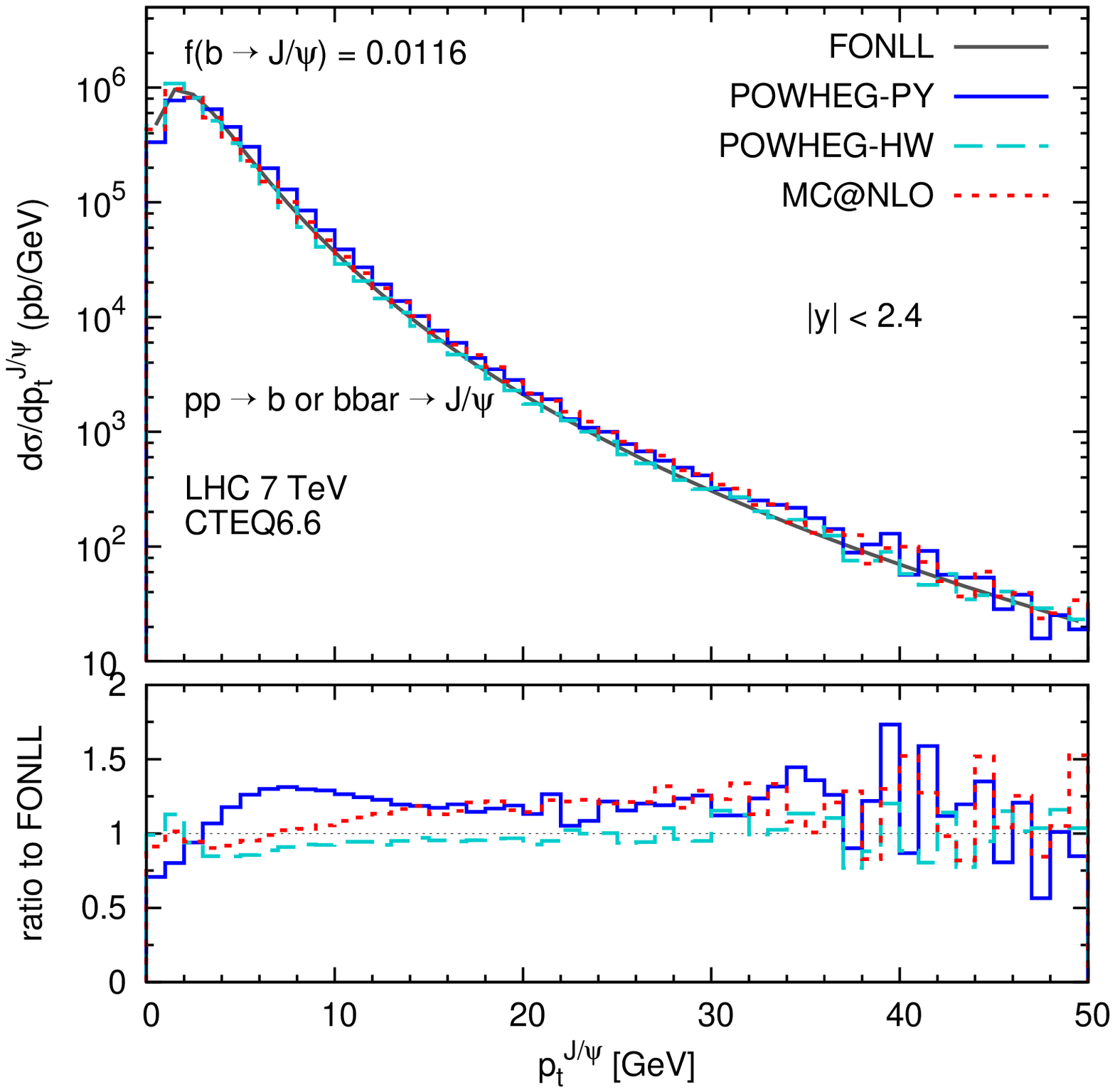}
\hfill
\includegraphics[height=0.48\textwidth,angle=-90]{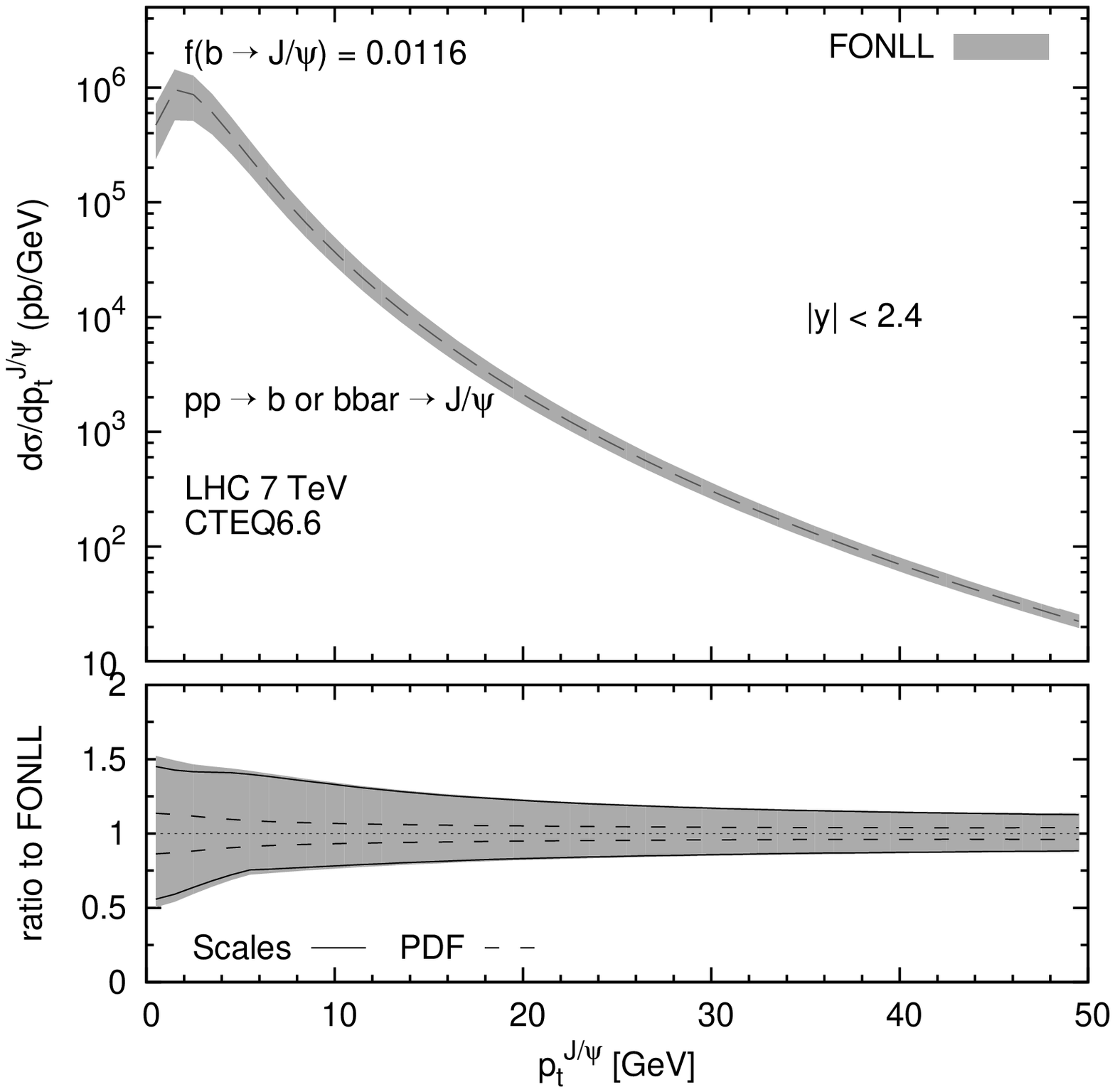}
\caption{\label{fig:jpsi}  Transverse momentum distribution of 
non-prompt $J\!/\!\psi$ in the central
rapidity region $|y|<2.4$. 
Left plot: comparison
  among the central predictions of our four benchmark calculations,
  \FONLL, \MCNLO, \PWHW\ and \PWPY. Right plot:
  theoretical systematics for the \FONLL\ calculation.
}
\end{center}
\end{figure}

Figure~\ref{fig:jpsi} compares predictions for the transverse momentum
distributions for the non-prompt  $J\!/\!\psi$ and $\psi(2S)$ production in the central rapidity
region $|y|<2.4$. In this case the prediction from the \PWPY\ implementation
differ quite markedly from the others, and data could provide discriminating
power. 

The first experimental measurements at 7 TeV from
CMS~\cite{Khachatryan:2010yr,Chatrchyan:2011kc}, LHCb~\cite{Aaij:2011jh} and
ATLAS~\cite{Aad:2011sp} have been compared to \FONLL\ predictions and generally found in
fairly good agreement. For example, LHCb~\cite{Aaij:2011jh}, after
separating the prompt and $b$-decay contributions, measures:
\be
\sigma^\mathrm{LHCb}(J\!/\!\psi~\mathrm{from}~H_b,  p_T < 14~\mathrm{GeV}, 2 < y < 4.5) = 1.14
\pm 0.01\pm 0.16~\mu b \\
\ee
to be compared with the \FONLL\ prediction 
(which includes a branching fraction BR$(b\to J\!/\!\psi$ = 0.0116)
\be
\sigma^\mathrm{FONLL}(J\!/\!\psi~\mathrm{from}~H_b, p_T < 14~\mathrm{GeV}, 2 < y < 4.5) = 
1.16~^{+0.55}_{-0.42}~\mu b \\
\ee
The \FONLL\ predictions describe also very well the J$\!/\!\psi$ $p_T$
spectrum, measured in the range 0--13~GeV, as shown in Fig.~9 of~\cite{Aaij:2011jh}.

More recently, ALICE~\cite{alice_psi_2012} reported:
\be
\sigma^\mathrm{ALICE}(J\!/\!\psi~\mathrm{from}~H_b,  p_T
>1.3~\mathrm{GeV}, \vert y\vert < 0.9) = 1.26
\pm 0.33~^{+0.23}_{-0.28}~\mu b \\
\ee
to be compared with the \FONLL\ prediction 
\be
\sigma^\mathrm{FONLL}(J\!/\!\psi~\mathrm{from}~H_b, p_T
>1.3~\mathrm{GeV}, \vert y \vert < 0.9) = 
1.33~^{+0.59}_{-0.48}~\mu b \\
\ee

A possible exception to the generally good agreement is the observation, 
made by CMS in \cite{Chatrchyan:2011kc}, that the experimental cross section for
non-prompt $J\!/\!\psi$ and $\psi(2S)$ production tends to fall 
off at large transverse momentum slightly faster than the \FONLL\ prediction.
More data at even larger $p_T$ will help clarify this issue.

\begin{figure}[t]
\begin{center}
\includegraphics[height=0.55\textwidth,angle=-90]{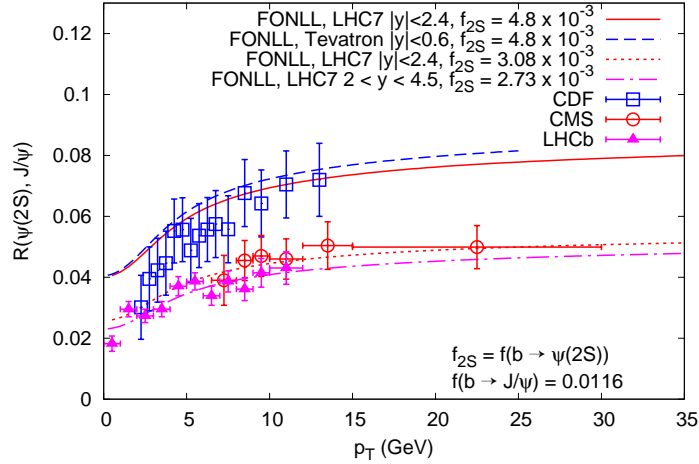}
\caption{\label{fig:psi2s-jpsi} \FONLL\ prediction of the ratio of the cross 
section for non-prompt (i.e. from $b$-hadrons) 
production of $\psi(2S)$ and $J\!/\!\psi$ (each multiplied by its own branching
ratio into muons) as a function of their transverse momentum, compared to
experimental data from CDF~\cite{Acosta:2004yw} at the Tevatron and
CMS~\cite{Chatrchyan:2011kc} and LHCb~\cite{LHCb-psi2s} at the LHC.} 
\end{center}
\end{figure}

In the same paper, CMS note that the measured non-prompt $\psi(2S)$ production is
quite smaller than the \FONLL\ prediction evaluated using the central value
of the branching fraction BR$(b\to\psi(2S)) = 4.8 \times
10^{-3}$~\cite{Nakamura:2010zzi}.
In order to better understand this potential discrepancy, it is useful to
consider the ratio
\be
R(\psi(2S), J\!/\!\psi) \equiv
\frac{
\frac{d\sigma}{dp_T}(b\to \psi(2S)) \mathrm{BR}(\psi(2S)\to \mu^+\mu^-)
}{
\frac{d\sigma}{dp_T}(b\to J\!/\!\psi) \mathrm{BR}(J\!/\!\psi\to \mu^+\mu^-)
}
\ee
as a function of the transverse momentum of the quarkonium.
Figure~\ref{fig:psi2s-jpsi} shows the \FONLL\ predictions for this ratio,
compared to data from CDF at the
Tevatron~\cite{Acosta:2004yw} and CMS~\cite{Chatrchyan:2011kc} and LHCb~\cite{LHCb-psi2s} at the LHC. One can 
see that the agreement with the
Tevatron data is acceptable when one uses BR$(b\to\psi(2S)) = (4.8 \pm 2.4 ) \times
10^{-3}$~\cite{Nakamura:2010zzi} and considers its large (O(50\%)) uncertainty, 
which is not shown in the plot.
The data from the CMS and the LHCb collaborations seem
instead  to be somewhat at variance with the CDF ones whereas, according to
the \FONLL\ prediction, no large difference should be present. One could still 
exploit the large uncertainty of BR$(b\to\psi(2S))$ to lower significantly the
theoretical prediction and make it compatible with all measurements within the
uncertainties.  In fact, CMS have used
their measurement of this ratio and the comparison to theoretical predictions to 
extract~\cite{Chatrchyan:2011kc} a  new value for the BR$(b\to\psi(2S))$ 
branching fraction,
\be
\mathrm{BR}(b\to\psi(2S))^\mathrm{CMS} = (3.08 \pm 0.12 \pm 0.13 \pm 0.42)\times 10^{-3} \, .
\ee

LHCb have more recently performed a similar extraction of 
$\mathrm{BR}(b\to\psi(2S))$ from their data, and have obtained
\be
\mathrm{BR}(b\to\psi(2S))^\mathrm{LHCb} = (2.73 \pm 0.06 \pm 0.16 \pm
0.24)\times 10^{-3} \, ,
\ee
fully compatible with the CMS determination.

When these new branching ratios are used in the \FONLL\ prediction, instead of
the $(4.8 \pm 2.4)\times 10^{-3}$ quoted above,
the results are, essentially by construction of course, in full agreement
with the LHC measurements, as shown in Fig.~\ref{fig:psi2s-jpsi}. On the other
hand, the older CDF data now appear to stand out, albeit being potentially marginally
compatible with the LHC ones within the respective experimental uncertainties. Since the
theoretical predictions for this ratio are largely insensitive to the specific experimental
setup, be it the center of mass energy (Tevatron or LHC) or even the rapidity acceptance cuts,
the discrepancy is likely of experimental origin.

\section{Open charm and bottom production from inclusive semileptonic decays}

In a fashion similar to that employed to describe $J\!/\!\psi$ and $\psi(2S)$
production from $b$-hadron decays, one can describe the production of a lepton
($e$ or $\mu$) originating from the electroweak decay of a charm or a bottom
hadron. As before, the decay spectrum and the branching fraction can be extracted
from experimental measurements. This approach has been used with \FONLL\ for the
first time in \cite{Cacciari:2005rk}, where the prediction for the
production rate of
electrons from charm and bottom in $pp$ collisions at RHIC was calculated. The
results were eventually found in good agreement with measurements by the
PHENIX~\cite{Adare:2006hc}
and the STAR\cite{Abelev:2006db} collaborations.

Three separate processes contribute to the final yield of leptons $\ell$ 
from a heavy hadron: $H_c\to \ell$, $H_b \to \ell$, and the secondary decay
$H_b \to H_c \to \ell$. The first contributes mainly at small transverse 
momentum
($p_T <$~5--10~GeV), the second dominates at larger $p_T$, while the secondary 
decay is largely negligible (see e.g. Fig. 3 of \cite{Cacciari:2005rk}).
All three processes have been modeled for \cite{Cacciari:2005rk},
and we use here exactly the same setup.

\begin{figure}[t]
\begin{center}
\includegraphics[height=0.48\textwidth,angle=-90]{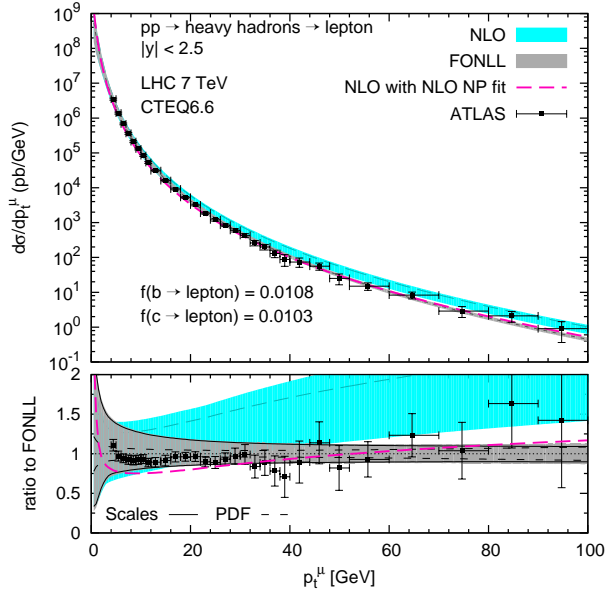}
\caption{\label{fig:leptons} Transverse momentum distribution of leptons from
heavy hadrons in the central
rapidity region $|y|<2.5$. The curves represent the sum of the three processes 
$H_c\to \ell$, $H_b \to \ell$ and $H_b \to H_c \to \ell$. No PDF uncertainties are
included in the NLO band. Data from~\cite{Aad:2011rr} are also shown.
Note that both muon charge states are included.}
\end{center}
\end{figure}

Figure \ref{fig:leptons} shows the cross section predicted by \FONLL\ for the sum of
the three processes, as they are largely indistinguishable experimentally. Two
NLO-level  predictions are also shown. Both make use of the fixed-order heavy quark
production NLO calculation for hadronic collisions. In one case (the 
upper band, blue in colour display, and
labeled simply `NLO') the same non-perturbative fragmentation functions also employed
in the \FONLL\ case are used. In the case of the magenta dashed curve instead, labeled
`NLO with NLO NP fit', the non-perturbative fragmentation functions have been fitted
to $e^+e^-$ LEP data for $B$ and $D^*$ production using a fixed next-to-leading order
calculation.
The \FONLL\ and the `NLO' predictions differ significantly at large $p_T$
when the same non-perturbative fragmentation functions are used, a consequence of the
resummation effects included in \FONLL\ . This difference is however
largely compensated -- as expected -- by the adjustment of the non-perturbative 
contribution, and one
can see that the magenta dashed curve `NLO with NLO NP fit' is indeed much closer to the
\FONLL\ prediction. It should be noted, however, that this prediction employing the 
NLO fit to LEP data is expected to be valid only in a
limited region, where transverse momenta are comparable to the scale $\mu$ at 
which the fit has been performed ($\mu=M_Z \simeq 90$~GeV).

Data for muon production are available from
ATLAS~\cite{Aad:2011rr} up to very large transverse momentum, and  one can see in
Fig.~\ref{fig:leptons} how they seem to be  better described by the resummed \FONLL\
prediction than by the fixed order one. Comparing to `NLO with NLO NP fit' rather than
to `NLO' may appear to largely wash out the edge of FONLL at large $p_T$, but the
agreement with the data seems to deteriorate at moderate $p_T$, as it may be expected
from the use of non-perturbative fragmentation functions fitted with a fixed order 
calculation
and at the much larger scale given by the $Z$ mass. In fact, the good agreement with the 
`NLO with NLO NP fit' is obtained by trading away the correct description of multiple
quasi-colliner emissions via resummation in exchange of their effective inclusion in the 
non-perturbative fragmentation functions fitted to the NLO calculation. 
While theoretically less ambitious, the 
latter procedure is also bound to fail when scales much different than $M_Z$ are probed.

\begin{figure}[t]
\begin{center}
\includegraphics[height=0.49\textwidth,angle=-90]{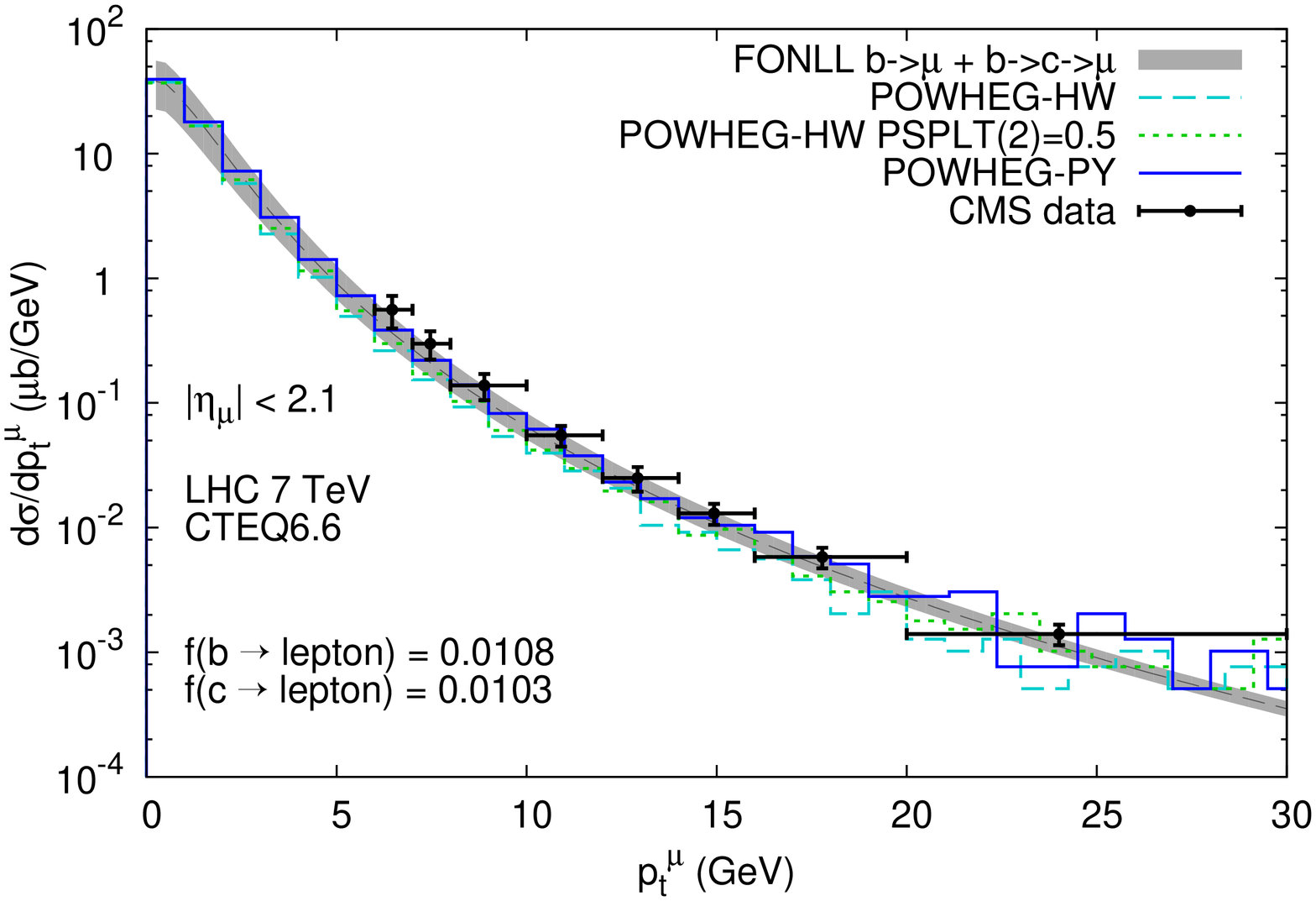}
\includegraphics[height=0.49\textwidth,angle=-90]{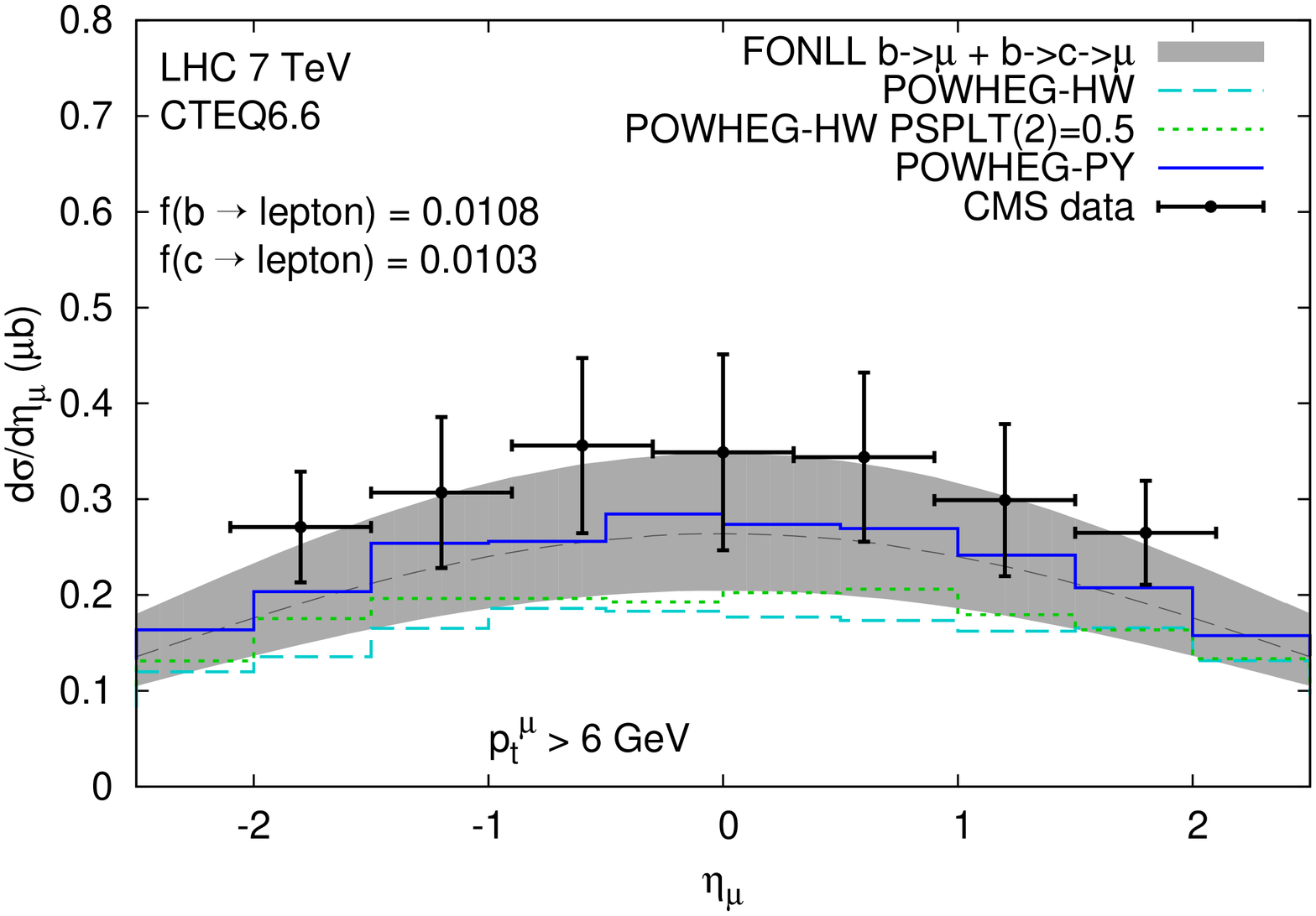}
\caption{\label{fig:cms-muons} CMS data for muon production from $b$-hadrons decays
compared to \FONLL\ and \PW\ predictions. Note that both charge states (i.e. muons
from both $b$ and $\bar b$) are included.}
\end{center}
\end{figure}

The ALICE Collaboration has also produced comparisons of distributions of 
leptons from heavy hadrons with \FONLL\ predictions, i.e. muons in the 
forward rapidity region $2.5 < y < 4$ (see for instance Fig.~3 of
\cite{Abelev:2012pi}) and electrons in the central region $|y| < 0.5$ (see
Fig.~11 of~\cite{Abelev:2012xe}): the overall agreement is similar to the one with the ATLAS
 data shown in Fig.~\ref{fig:leptons}.

CMS also published results for transverse momentum and rapidity distributions for muons
from $b$-hadrons~\cite{Khachatryan:2011hf}. The plots in
Fig.~\ref{fig:cms-muons} compare their results
with \FONLL\ and \PW\ predictions, and are found to be compatible within uncertainties. From
their measurements CMS extract a total visible cross section for muons from
$b$-hadron decays,
\be
\sigma^\mathrm{CMS}(pp\to H_b+X\to\mu+X',~p_T^\mu > 6~\mathrm{GeV},~|y^\mu| < 2.1) = 1.32 \pm 0.01 \pm 0.3 \pm 0.15~\mu b
\ee
to be compared with the \FONLL\ prediction (which uses the branching ratios 
BR$(b\to\ell) = 0.0108$ and BR$(b\to c\to\ell) = 0.096$ \cite{Nakamura:2010zzi}) 
\be
\sigma^\mathrm{FONLL}(pp\to H_b+X\to\mu+X',~p_T^\mu > 6~\mathrm{GeV},~|y^\mu| < 2.1) = 0.855~^{+0.28}_{-0.19}~\mu b
\ee
Note that both charge states are included in these results.

\begin{table}[ht]
\begin{center}
\begin{tabular}{| l | l | l | l | l |}
\hline
Expt & Observable ($p_T$ in GeV) & $\sigma^\mathrm{exp}$ & $\sigma^\mathrm{\scriptscriptstyle{FONLL}}$ & Comments
\\
\hline {\small 
1: LHCb~\cite{Aaij:2010gn}} &
{\small $\sigma(H_b, 2\le \eta\le 6)$ 
} & {\small  
$75.3 \pm 11.4$~$\mu$b
} & {\small 
$70.8~^{+33.3}_{-24.4}$~$\mu$b
} & {\small 
average $b+\bar{b}$ 
} \\
\hline {\small

2: LHCb~\cite{Aaij:2012jd}} & {\small 
$\sigma( B^\pm, p_T < 40, 2 < y < 4.5)$
} & {\small  
$41.4 \pm 3.4$~$\mu$b
} & {\small 
$40.1~^{+19.0}_{-14.5}$~$\mu$b
} & {\small 
$f(b\to B^{-})$ = 0.403 
} \\
\hline {\small

3: CMS~\cite{Chatrchyan:2011pw}} & {\small 
$\sigma( B^0,~p_T^B > 5,~|y^B| < 2.2)$
} & {\small  
$33.2 \pm 4.3$~$\mu$b
} & {\small 
$25.5~^{+10.5}_{-7.1}$~$\mu$b
} & {\small 
$f(b\to B^{0})$ = 0.403
} \\
\hline {\small

4: CMS~\cite{Khachatryan:2011mk}} & {\small 
$\sigma( B^+,~p_T^B > 5,~|y^B| < 2.4)$
} & {\small  
$28.1 \pm 4.4$~$\mu$b
} & {\small 
$27.2~^{+11.2}_{-7.5}$~$\mu$b
} & {\small 
$f(b\to B^{-})$ = 0.403
} \\
\hline {\small

5: CMS~\cite{Chatrchyan:2011vh}} & {\small 
$\sigma(B^0_s,~8 < p_T^B < 50,~|y^B| < 2.4)$
} & {\small 
$6.9 \pm 0.8$~nb
} & {\small 
$4.5~^{+2.3}_{-1.9}$~nb
} & {\small  
$f(b\to B^{0}_s)$ = 0.11
} \\
& {\small 
$\times \mathrm{BR}(B^0_s\to J\!/\!\psi\,\phi)$ }
&

&
{\small (includes BR}
& {\small  
BR$(B^0_s\to J\!/\!\psi\,\phi) =$
} 
\\
& & & {\small uncertainty)}& {\small  
~~~~ $(1.4 \pm 0.5)\times
10^{-3}$
} 
\\

\hline {\small

6: LHCb~\cite{Aaij:2011jh}} & {\small 
$\sigma( H_b \to J\!/\!\psi,  p_T^{\psi} <
14, 2 < y_{\psi} < 4.5)$
} & {\small  
$1.14 \pm 0.16$~$\mu$b
} & {\small 
$1.16~^{+0.55}_{-0.42}$~$\mu$b
} & {\small 
BR$(b\to J\!/\!\psi)$ = 0.0116
} \\
\hline {\small

7: ALICE~\cite{alice_psi_2012}} & {\small 
$\sigma( H_b \to J\!/\!\psi,  p_T^{\psi} >
1.3, \vert y_{\psi}\vert  < 0.9)$
} & {\small  
$1.26 \pm 0.16$~$\mu$b
} & {\small 
$1.33~^{+0.59}_{-0.48}$~$\mu$b
} & {\small 
BR$(b\to J\!/\!\psi)$ = 0.0116
} \\
\hline {\small

8: CMS~\cite{Khachatryan:2011hf}} & {\small 
$\sigma( H_b\to\mu,~p_T^\mu > 6,~|y^\mu| < 2.1)$
} & {\small  
$1.32 \pm 0.34$~$\mu$b
} & {\small 
$0.855~^{+0.28}_{-0.19}$~$\mu$b
} & {\small 
BR$(b\to\ell) = 0.0108$
}
\\
& & & & {\small 
BR$(b\to c\to\ell) = 0.096$
} \\
\hline

\end{tabular}
\caption{\label{tab:summary} Summary of various cross-section measurements,
  compared against the \FONLL\ predictions. The numbers labeling each
  measurement refer to the entries in Fig.~\ref{fig:summary}.}
\end{center}
\end{table}

\begin{figure}[t]
\begin{center}
\includegraphics[width=0.5\textwidth,angle=90]{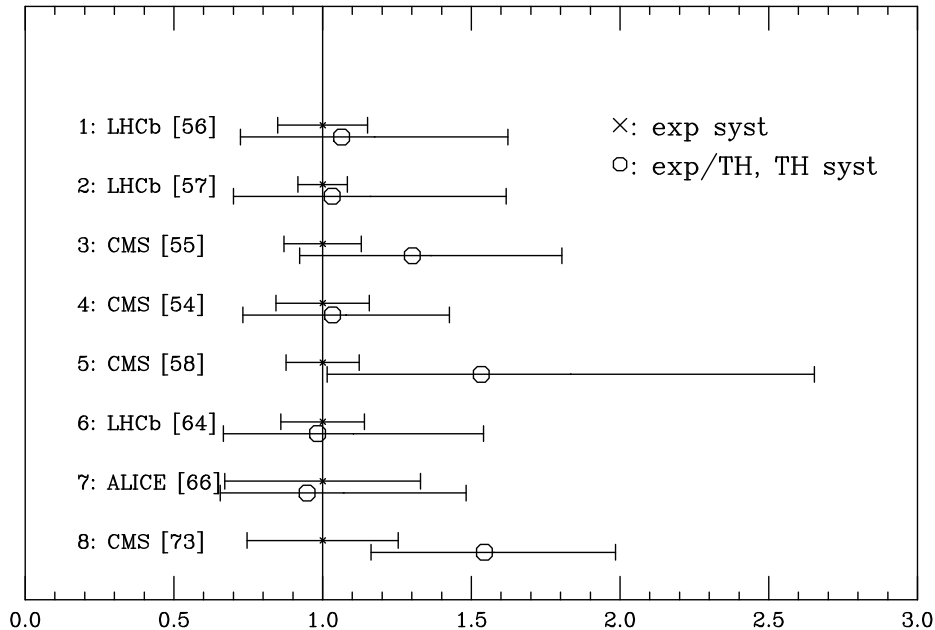}
\caption{\label{fig:summary} Graphical display of the cross section
  results listed in Table~\ref{tab:summary}. We show the ratios of
  experimental data over the \FONLL\ calculations, with the sizes of
  experimental and theoretical systematics. The numbers labeling each
  measurement refer to the entries in Table~\ref{tab:summary}.
}
\end{center}
\end{figure}

\section{Conclusions}
We have presented theoretical predictions for heavy quark cross
sections for various observable final states within realistic
acceptance cuts. In a number of cases, we compared the predictions of
\FONLL, \MCNLO\ and \PW\ with \HW\ and/or \PY. These predictions
are also compared directly with experimental measurements at the LHC,
and this paper is meant to provide a detailed description and a record
of these theoretical results.

The curves presented in this article, as well as others directly
delivered to the experimental collaborations, were mostly obtained
before any data were available.  They have not in the least been
influenced by the LHC measurements, and are a direct extension to LHC
energy of a framework previously built and validated using LEP and
Tevatron data.

One can note that, while \FONLL, \MCNLO\ and \PW-based predictions
generally agree with each other 
in the moderate transverse momentum regions, some
differences can be observed at large $p_T$. The small-$p_T$
agreement is expected, as by construction all three frameworks must
return the NLO cross section when integrated over the whole phase
space, and this result will be dominated by the small-$p_T$ region. At
large $p_T$, instead, differences can arise. Part of them may be due
to the Monte Carlo based approaches only resumming the quasi-collinear
logarithms to `almost' leading log accuracy rather than
next-to-leading one, but this is expected to be a minor effect. More
importantly, non-perturbative fragmentation contributions that had
been adjusted to data in \PY\ and \HW\ are likely not fully
appropriate to be used in a matched context where hard gluon radiation
to next-to-leading order level is also included. These
non-perturbative parametrisations should therefore be re-adjusted to
$e^+e^-$ data within the same \MCNLO\ and \PW\ frameworks that are
then going be used to produce predictions for hadronic collisions,
analogously to what is done with \FONLL. Lack of an appropriate tuning
of the non-perturbative hadronization of heavy quarks could provide an
explanation for the discrepancy between the fragmentation function of
$D^*$ mesons inside jets, measured in Ref.~\cite{Aad:2011td}, and the
theoretical predictions, based on NLO PSMCs, shown in that paper.

The agreement between the LHC experimental data we discussed and
theoretical predictions is otherwise generally very good. 
We summarize the various cross sections reported in this paper in
Table~\ref{tab:summary} and in Fig.~\ref{fig:summary}.
The comparison of theory and data at small $p_T$, as well as at large
rapidity, does not point to the existence of a new dynamical regime for
the production of heavy quarks (e.g. one dominated by small-$x$
effects), beyond what can be accounted for by fixed-order NLO
calculations. The measurements at large transverse momentum,
furthermore, seem to favour a description based on FONLL, underscoring
the importance of resumming large logarithms of $p_T^Q/m_Q$. 

Other sets of measurements not discussed in this paper, such as the
$E_T$ spectrum and angular correlations of jets containing $b$
quarks~\cite{ATLAS:2011ac,Chatrchyan:2012dk}, are consistent with
these findings. Some discrepancies with
\MCNLO, on the other hand, have been reported recently in a study of
angular correlations between $b$-hadron
pairs~\cite{Khachatryan:2011wq}.

It is important to consider that, while the overall
theoretical systematics is typically large, most of it is highly
correlated in different measurements and in different kinematic
regions. This is certainly the case of mass, fragmentation and 
PDF systematics, but
it is also true of the scale systematics.  When these correlations are
taken into account, the agreement with the experimental data is even more
remarkable, particularly considering the efforts that were required,
during the first 10 years of measurements at the Tevatron, to
reach such a level of consistency.

\section*{Acknowledgements}

We acknowledge numerous interactions with many experimental colleagues who, before
and during their analyses, suggested the appropriate final states and acceptance
cuts for the production of these predictions. In particular, we wish to thank
Vincenzo Chiochia, Zaida Conesa del Valle, Andrea Dainese, Biagio Di Micco,
Guenther Dissertori, Daniel Froidevaux, Leonid Gladilin, Jibo He, Vato
Kartvelishvili, Aafke Kraan,  Rolf Oldeman, Fabrizio Palla, Gabriella Pasztor,
Fabrizio Petrucci, Patrick Robbe, Giovanni Sabatino, Michael Schmelling, Enrico
Scomparin, Sheldon Stone. We also thank the Galileo Galilei Institute for
Theoretical Physics for the hospitality and the INFN for partial support 
while part of this work was performed.

%\appendix

\end{document}